\documentclass[useAMS,usenatbib,manuscript]{mn2e}
\usepackage{graphicx}
\usepackage{amsmath} 
\usepackage{amssymb} 
\usepackage{natbib}

\title[Dwarf galaxies in voids]
{Dwarf galaxies in voids: \\ Suppressing star formation with photo-heating}

\author[Hoeft et al.]
   {
  Matthias Hoeft$^1$, Gustavo Yepes$^2$, Stefan Gottl\"ober$^3$, and Volker Springel$^4$ \\
  $^1$International University Bremen, Campus Ring 1, 28759 Bremen, Germany \\
  $^2$Grupo de Astrofisica, Universidad Autonoma de Madrid, Cantoblanco, 28039 Madrid, Spain \\
  $^3$Astrophysikalisches Institut Potsdam, An der Sternwarte 16, 14482 Potsdam, Germany\\
  $^4$Max-Planck-Institut f\"ur Astrophysik, Karl-Schwarzschild-Str. 1, Garching bei M\"unchen, Germany
  }
\date{}

\def\LaTeX{L\kern-.36em\raise.3ex\hbox{a}\kern-.15em
    T\kern-.1667em\lower.7ex\hbox{E}\kern-.125emX}

\newcommand{\Unitkmpers}{\:\mathrm{km/s}}

\def\hMpc{\ifmmode{\:h^{-1}\,{\rm Mpc}}\else{$h^{-1}\,{\rm Mpc}$}\fi}
\def\hkpc{\ifmmode{h^{-1} \,{\rm kpc}}\else{$h^{-1}\,{\rm kpc}$}\fi}
\def\hMsun{\ifmmode{\:h^{-1}\,M_\odot}\else{$h^{-1}\,M_\odot$}\fi}
\newcommand{\kms}{\Unitkmpers}

\newcommand{\G}{{\sc Gadget}}
\newcommand{\g}{\G\ }

\newcommand{\Msunh}{\,\mbox{$h^{-1}\,{\rm M}_\odot$}}
\def\hMsun{\ifmmode{h^{-1}{\rm M}_\odot}\else{$h^{-1}M_\odot$}\fi}

\begin{document}

\renewcommand{\topfraction}{0.6}
\renewcommand{\textfraction}{0.3}

\label{firstpage}

\maketitle

\begin{abstract}
  
  We study structure formation in cosmological void regions using
  high-resolution hydrodynamical simulations. Despite being
  significantly underdense, voids are populated abundantly with small
  dark matter halos which should appear as dwarf galaxies if their
  star formation is not suppressed significantly. We here investigate
  to which extent the cosmological UV-background reduces the baryon 
  content of dwarf galaxies, and thereby limits their
  cooling and star formation rates.  Assuming a Haardt \& Madau
  UV-background with reionisation at redshift $z=6$, our samples of
  simulated galaxies show that halos with masses below a
  characteristic mass of $M_c(z=0) = 6.5 \times 10^9 \: h^{-1} \, {\rm
  M}_\odot$ are baryon-poor, but in general not completely empty,
  because baryons that are in the condensed cold phase or are already
  locked up in stars resist evaporation. In halos with mass $M
  \lesssim M_c$, we find that photo-heating suppresses further cooling
  of gas.  The redshift and UV-background dependent characteristic
  mass $M_c(z)$ can be understood from the equilibrium temperature
  between heating and cooling at a characteristic overdensity of
  $\delta \simeq 1000$. If a halo is massive enough to compress gas to
  this density despite the presence of UV-background radiation, gas is free
  to `enter' the condensed phase and cooling continues in the halo,
  otherwise it stalls.  By analysing the mass accretion histories of
  dwarf galaxies in voids, we show that they can build up a
  significant amount of condensed mass at early times before the epoch
  of reionisation.  Later on, the amount of mass in this phase remains
  roughly constant, but the masses of the dark matter halos continue
  to increase.  Consequently, photo-heating leads to a reduced baryon
  fraction in void dwarf galaxies, endows them with a rather old
  stellar population, but still allows late star formation to some
  extent.  We estimate the resulting stellar mass function for void
  galaxies. While the number of galaxies at the faint end is
  significantly reduced due to photo-heating, additional physical
  feedback processes may be required to explain the apparent paucity
  of dwarfs in observations of voids.
\end{abstract}

\begin{keywords}
cosmology: theory --
methods: numerical --
galaxies: evolution --
galaxies: formation
\end{keywords}

\section{Introduction}

Large regions of space that contain few or no galaxies can be clearly
identified in modern spectroscopic redshift surveys. About 25 years ago, such
`voids' have been first discovered \citep{gregory:78,joeveer:78,kirshner:81}, 
but it remains a challenge to explain why they are apparently
so empty. 

It is well known that hierarchical models of structure formation in
standard cold dark matter (CDM) cosmologies produce large underdense
regions in the distribution of matter
\citep{peebles:82,hoffman:82,weygaert:93}, but these underdense
regions still contain structural elements and bound halos, even though
the characteristic masses of these objects are several orders of
magnitude smaller than corresponding ones found in average regions of
the universe.  \citet{gottloeber:03} predicted that a typical
$20\:h^{-1}\,{\rm Mpc}$ diameter void should contain up to 1000 halos
with mass $\sim 10^9 \: h^{-1} \, {\rm M}_\odot$ and still about 50
halos with mass $\sim 10^{10} \: h^{-1} \, {\rm M}_\odot$. Assuming a
magnitude of $M_B = -16.5$ for the galaxy hosted by a halo of mass
$3.6 \times 10^{10}\hMsun$ \citep{mathis:02} predict that about
five such galaxies should be found in the inner regions of a typical
void of diameter $20 \hMpc$.

Over the last decade, there were many attempts to find dwarf galaxies in voids
\citep{lindner:96,popescu:97,kuhn:97,grogin:99}. An overall conclusion from
these studies has been that faint galaxies do not tend to fill up the voids
outlined by the bright galaxies. \citet{peebles:01} pointed out that the dwarf
galaxies in the Optical Redshift Survey (ORS) follow the distribution of
bright galaxies remarkably closely. Using the SDSS data release 2,
\citet{goldberg:05} measured the mass function of galaxies that reside in
underdense regions. They selected galaxies as void members if they had less
than three neighbours in a sphere with radius $7 \hMpc$. More than one
thousand galaxies passed their selection criterion (which differs slightly
from the criterion used in the numerical simulations of \citet{gottloeber:03}
and in this paper).  Their measurements are consistent with the predictions
from the numerical simulations once the tendency of more massive halos to
concentrate at the outer parts of voids (where they still may pass the
nearest neighbour selection criterion) is taken into account. However, the
observational situation is unclear for halo masses smaller than $\sim
10^{10}\hMsun$.  At present, there are no observational hints that a huge
number of dwarf galaxies in voids may exist, despite the large number of small
halos predicted by CDM models.

This finding resembles the `substructure problem' in galactic halos
\citep{klypin:99b,moore:99}.  A solution for both problems could arise
from physical processes capable of suppressing star formation in dwarf
galaxies.  The two major effects proposed in this context are
supernova feedback and heating of the gas in halos by the
UV-background radiation. The latter increases the thermal pressure,
and as a result, the gas in systems with $T_{\rm vir} \lesssim
10^4-10^5\:{\rm K}$ can be evaporated out of halos
\citep{umemura:84,dekel:87,efstathiou:92,babul:92}. Similarly,
supernova feedback could drive a galactic outflow that removes a
significant fraction of the gas in a dwarf galaxy
\citep{dekel:86,couchman:86}. However, the efficiency of supernova
driven winds depends strongly on the details of model assumptions
\citep{navarro:97,navarro:00,springel:03}. A general result is that
low-mass systems are much more easily affected by supernova driven
winds than larger ones \citep{maclow:99,navarro:00}.

During the epoch of reionisation, the gas temperature is raised to a few times
$10^4\:{\rm K}$. \citet{rees:86} argued that in dark matter halos with virial
velocities around $\sim 30 \: {\rm km \, s^{-1}}$ gas can then be confined in
a stable fashion, neither able to escape nor able to settle to the centre by
cooling.  The gas in significantly smaller systems is thought to be virtually
evaporated \citep{thoul:96,barkana:99}. However, due to central self-shielding
the evaporation of gas that has already cooled can be significantly delayed
\citep{susa:04,susa:04b}. Using radiative transfer simulations,
\cite{susa:04b} pointed out that halos may contain a significant amount of
stellar mass produced before reionisation occurred, even if their remaining gas
mass is evaporated during reionisation.

In a 3D Eulerian adaptive mesh refinement simulation,
\citet{tassis:03} found that the global star formation rate was
significantly reduced during the reionisation epoch. Their result also
indicated that stellar feedback enhances this effect
dramatically. Thus, an imprint of the epoch of reionisation may be
expected for the stellar population in dwarf galaxies. Indeed, almost
all dwarf galaxies appear to have an early epoch of star formation
\citep{mateo:98}, but there is no distinct time at which star
formation becomes generally suppressed \citep{grebel:04}.  

In this paper, we use high-resolution hydrodynamical simulations of
cosmological void regions to analyse the star formation and cooling processes
of void galaxies. In particular, the simulations are well suited for studying
the evolution of isolated dwarf galaxies from the epoch of reionisation to
the present. This allows us to examine whether the UV background has a
sufficiently strong effect on dwarf galaxies, keeping them faint enough such
that their abundance can be reconciled with observations. We also determine
the characteristic mass scale below which cooling is suppressed by the UV
background. An analysis of the spatial distribution of dwarfs, the impact of
supernova feedback, and the spectral properties of the stellar content of the
formed dwarf galaxies will be discussed separately.

Our study is organised as follows.  Details of our simulations are described
in Section~\ref{sec-simu}.  In Section~\ref{sec-galax}, we analyse first the
baryon fraction as a function of both halo mass and redshift. Then we
investigate the mass growth of the condensed phase which consists of both 
stars and cold dense baryons. We identify the characteristic mass below which 
halos are subject to evaporation. Finally, we estimate the galaxy mass function 
in void regions. We discuss and summarise our results in Section~\ref{sec-discuss}.

\begin{table*}
\begin{tabular}{l c c c r c c c c c }
\hline
{Simulation}& Refinement & $M_{\rm gas}$ & $M_{\rm dark}$  & Particle &
Star &  \multicolumn{4}{c}{Feedback parameters}  \\
{name} &  levels & $(10^6 \Msunh)$ & $(10^6 \Msunh)$
 & number & fraction & $\beta $ &  A & $T_{SN}$ & $J^{\rm UV}_0$ \\

\hline

void2         & 3 & 5.51 & 34.2 &  5,068,359  & 0.060 & 0.1 & 1000 & $10^8$ &  0.95 \\
basic         & 3 & 1.50 & 8.24 &  7,376,094  & 0.048 &  '' &  ''  & ''     &  '' \\
high-res      & 4 & 0.18 & 1.03 & 43,544,537  & 0.053 &  '' &  ''  & ''     &  '' \\
high-UV       & 3 & 1.50 & 8.24 &  7,390,626  & 0.024 &  '' &  ''  & ''     &  95.0 \\
low-UV        & 3 &  ''  &   '' &  7,676,786  & 0.079 &  '' &  ''  & ''     & 0.0095 \\
no-UV         & 3 &  ''  &   '' &  7,873,866  & 0.118 &  '' &  ''  & ''     & 0.0 \\
imp-heat-15   & 3 &  ''  &   '' &  7,504,500  & 0.046 &  '' &  ''  & ''     & 0.95 \\
imp-heat-50   & 3 &  ''  &   '' &  7,499,772  & 0.044 &  '' &  ''  & ''     & 0.95 \\
no-feedback   & 3 &  ''  &   '' &  7,384,373  & 0.047 &  0  &  0   &  0     & 0.95 \\
\hline 

\end{tabular}

\caption{ Main characteristics of the void simulations. $M_{\rm gas}$ and
$M_{\rm dark}$ denote the mass of a gas and of a dark matter particle in the
simulation, respectively. Feedback parameters are according to model described
in \citet{springel:03}. The UV-flux, $J^{\rm UV}_0$, at $z=0$ is
given in units of $10^{-23}\:{\rm ergs \, s^{-1} \, cm^{-2} \, sr^{-1} \,
Hz^{-1}}$. For the description of the simulation with an additional heat pulse see Sec.~\ref{sec-baryon-frac}.
\label{table1}
}

\end{table*}

\section{Simulations}

\label{sec-simu}

\subsection{Numerical method}

Our simulations have been run with an updated version of the parallel Tree-SPH
code \g \citep{springel:01}. The code uses an entropy-conserving formulation
of SPH \citep{springel:02} which alleviates problems due to numerical over-cooling.  The
code also employs a new algorithm based on the Tree-PM method for the $N$-body
calculations which speeds up the gravitational force computation significantly
compared with a pure tree algorithm.

\begin{figure}
  \begin{center}
  \includegraphics[width=0.45\textwidth,angle=0]{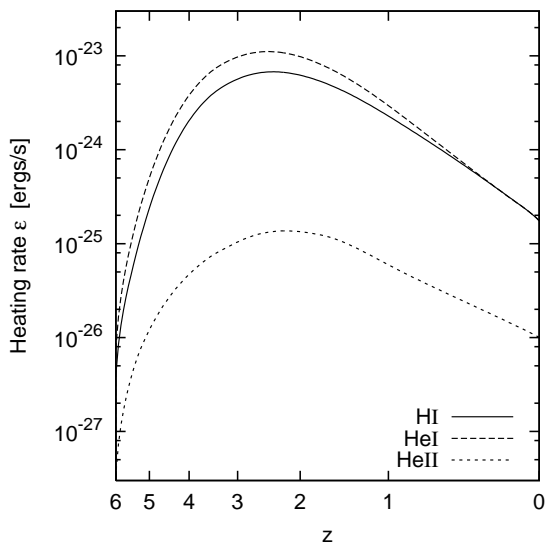}
  \end{center}
  \caption{
      Photo-heating rates due to the ambient UV-background as a
      function of redshift. We adopt a time evolution for the UV-background as
      given by \citet{haardt:96}.  Reionisation in this model takes place at
      $z=6$.  }
  \label{fig-heating-rates}
\end{figure}

Radiative cooling processes for an optically thin primordial mix of helium and
hydrogen are included, as well as photo-ionisation by an external, spatially
uniform UV-background. The net change of thermal energy content is
  calculated by closely following the procedure of \citet{katz:96} for solving
  the rate equations. We also use the set of cross sections cited in this
  paper, but adopt a slightly modified \citet{haardt:96} UV-background with
  heating rates as depicted in Fig.~\ref{fig-heating-rates}.

\begin{figure*}
  \begin{center}
  \includegraphics[width=0.3\textwidth,angle=-90]{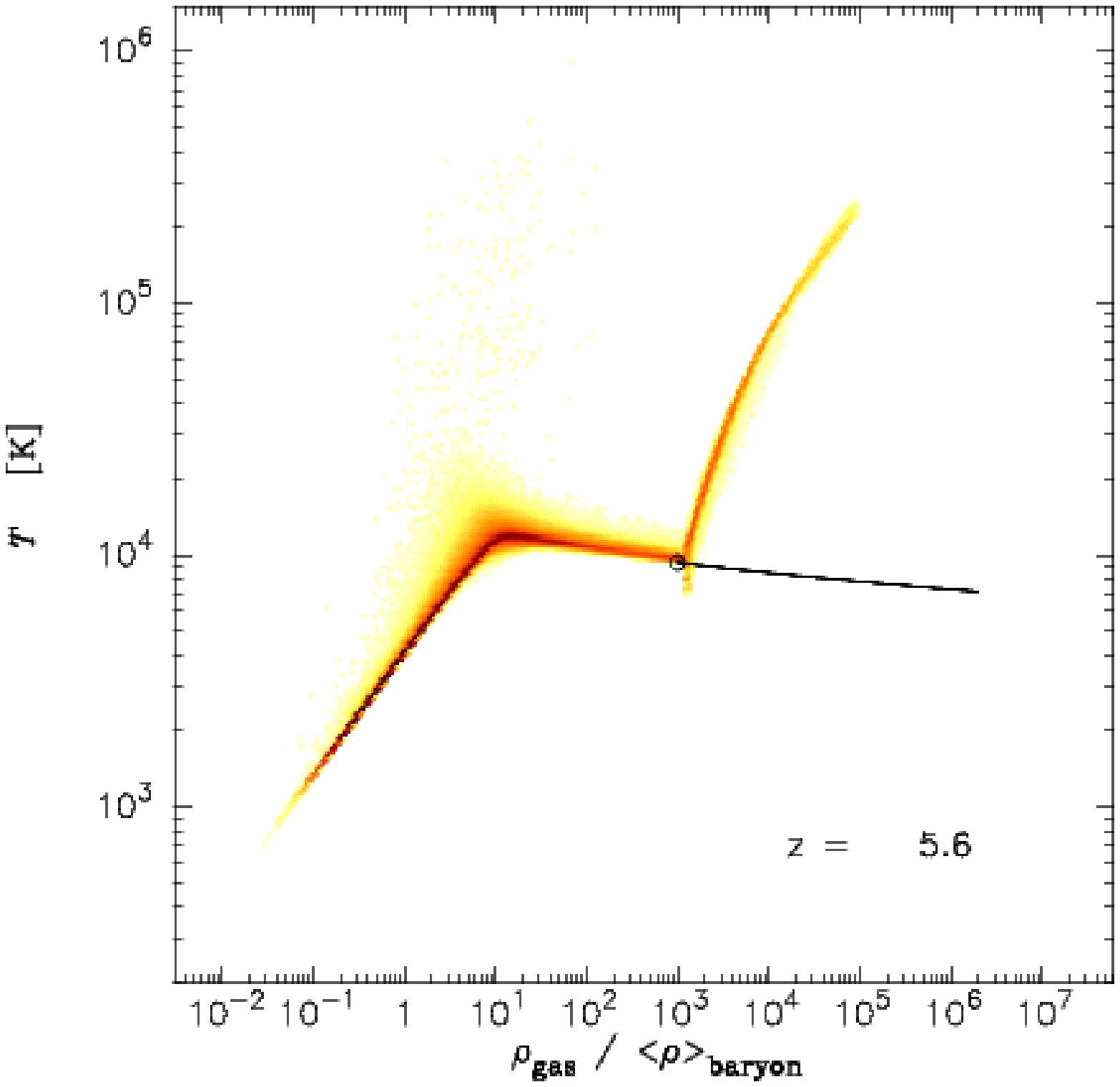}
       \hfill
  \includegraphics[width=0.3\textwidth,angle=-90]{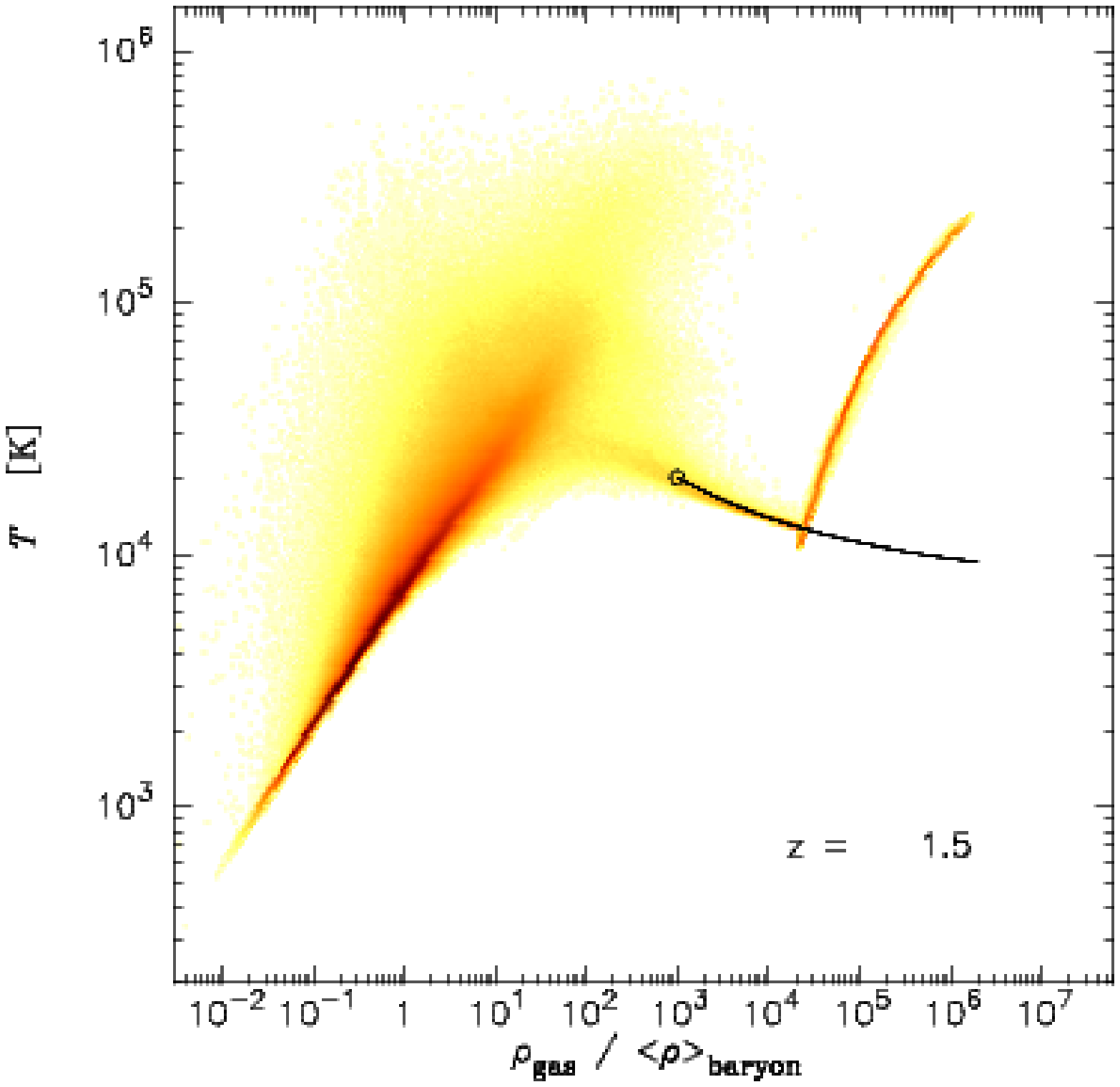}
       \hfill
  \includegraphics[width=0.3\textwidth,angle=-90]{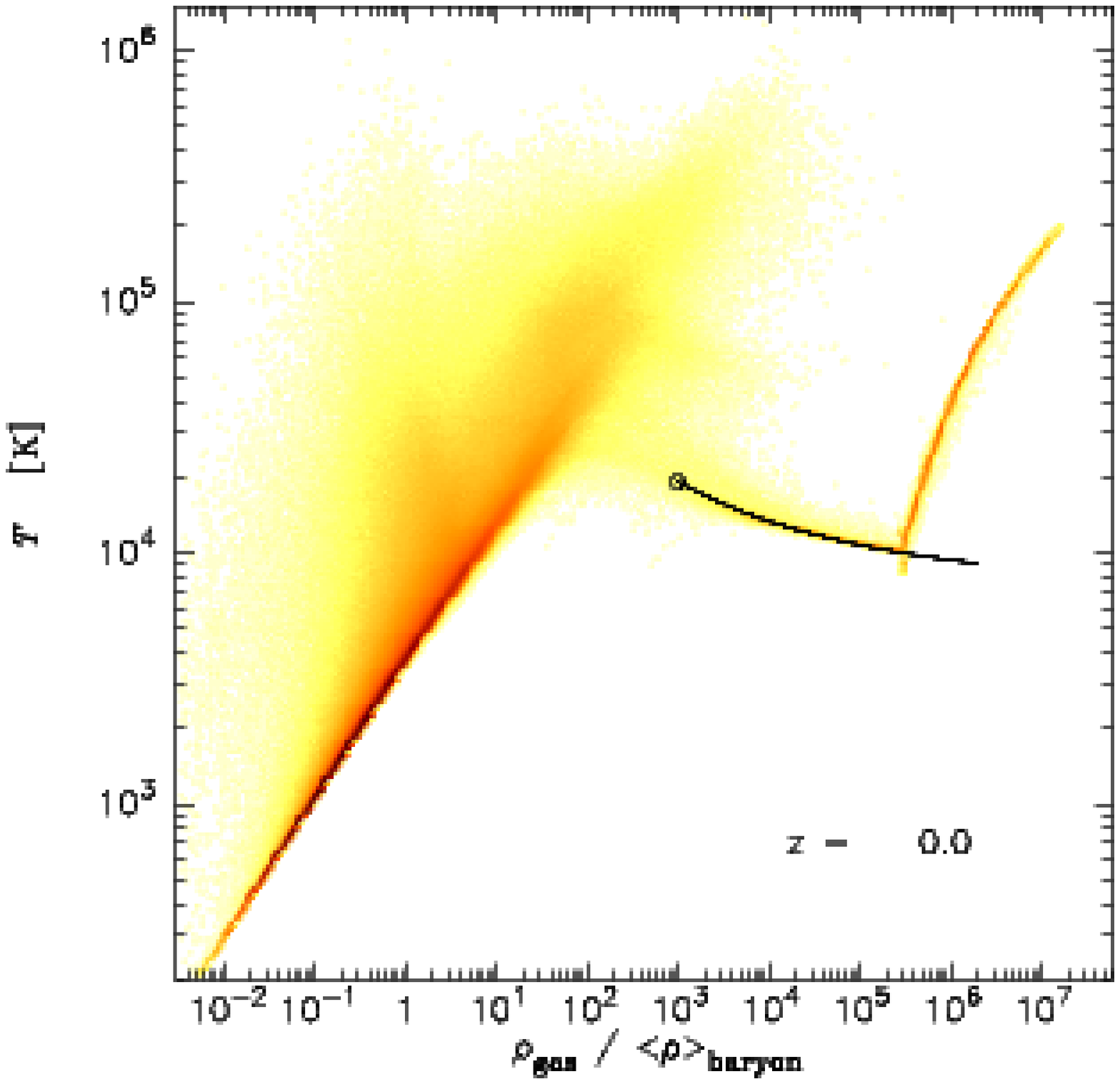}
  \end{center}
  \caption{ 
The distribution of particles in the density -
temperature phase diagram for different redshifts for the high-res
run. The solid curves give the equilibrium line. It is derived from the
heating-cooling module in the code by computing the temperature at
which the heating rate equals the net cooling rate. We assume that the
`entry' into the condensed branch occurs at $10^3\:\langle \rho_{\rm
baryon}\rangle$. The circles indicate the derived entry temperatures
for the individual redshifts. See Sec.~\ref{sec-how-to-supp} and Fig.~\ref{fig-J-Teq}
for a discussion of the equilibrium and entry temperatures.  }
  \label{fig-rho_T_phase_diagram}
\end{figure*}

The physics of star formation is treated in the code by means of a
sub-resolution model in which the gas of the interstellar medium (ISM) is
described as a multiphase medium of hot and cold gas
\citep{yepes:97,springel:03}. Cold gas clouds are generated due to cooling and
are the material out of which stars can be formed in regions that are
sufficiently dense.  Supernova feedback heats the hot phase of the ISM and
evaporates cold clouds, thereby establishing a self-regulation cycle for star
formation. The heat input due the supernovae also leads to a net
pressurisation of the ISM, such that its effective equation of state becomes
stiffer than isothermal, see Fig.~\ref{fig-rho_T_phase_diagram}.  
This stabilises the dense star forming gas in
galaxies against further gravitational collapse, and allows converged
numerical results for star formation even at moderate resolution.  We also
follow chemical enrichment associated with star formation, but we have
neglected metal-line cooling in computing the cooling function.  See
\citet{springel:03} for a more detailed description of the star formation
model implemented in the \g code.

\subsection {Initial conditions}

Using the mass refinement technique described by
\citet{klypin:01}, we simulate void regions with high mass
resolution, embedded in a proper cosmological environment.  Our voids
have been selected for resimulation from two periodic computational
boxes of side-lengths $L= 80\:h^{-1}\,{\rm Mpc}$ and $50\:h^{-1}\,{\rm
Mpc}$, respectively. To construct suitable initial conditions, we
first created an unconstrained random realization at very high
resolution, using the $\Lambda$CDM power spectrum of perturbations. For
the large box, $N=1024^3$ particles were used, while for the smaller
box, we employed $2048^3$ ($\sim 8.6$ billion) particles.  The initial
displacements and velocities of the particles were calculated using
all waves ranging from the fundamental mode $k=2\pi/L$ to the Nyquist
frequency $k_{\rm ny}=2\pi/L\times N^{1/3}/2$. To produce initial
conditions at lower resolution than this basic high-resolution
particle set-up, we then merged particles, assigning to merged
particles a velocity and a displacement equal to the average values of
the original small-mass particles.

In this way, we first run $128^3$ low-resolution simulations until the
present epoch and selected the void regions from them. The algorithm
for identifying the voids is described in detail in
\citet{gottloeber:03}. It allows us to select void regions with
arbitrary shape. To this end, the method starts from a spherical
representation of the void which is then extended by spheres of
smaller radius, which are added from the surface of the void into all
possible directions. However, in the present application we have
restricted the resimulation to a spherical void region to avoid
ambiguities in the definition of allowed deviations from spherical
shape.

In the second simulation step we use the original sample of small-mass
particles in the regions of interest when we construct initial conditions.
Thus we reach a mass resolution within the void regions that corresponds to
the $1024^3$ or $2048^3$ set-up, respectively. We use a series of shells
around the voids where we progressively merge more and more of the particles
until the effective resolution of $128^3$ particles is reached again far away
from the voids. This procedure ensures that the voids evolve in the proper
cosmological environment and with the right gravitational tidal fields.
Mixing of particles of different mass occurs only in the shells surrounding
the high resolution voids. Finally, we split the particles in the regions of
high mass resolution into dark matter and gas particles.  For all simulations,
we adopted a concordance cosmological model with $\Omega_m = 0.3$,
$\Omega_\Lambda = 0.7$, $\Omega_b = 0.04$, $h=H_0 / (100 \, {\rm km \, s^{-1}
  \, Mpc^{-1}} ) = 0.7$ and $\sigma_8 = 0.9$.

\subsection{Simulation runs}

Using the multi-mass technique described above, we have re-simulated a void
region in the $80\:h^{-1}\,{\rm Mpc}$ box with three levels of refinement.
The mass of a dark matter particle in the void is $3.4 \times 10^7\Msunh$. The
corresponding SPH gas particles have an initial mass of $5.5 \times10^6
\hMsun$. Note that some gas particles may reduce their mass during the run (or
vanish entirely) if they undergo star formation and create new collisionless
star particles. In our analysis of the simulation results, we in general only
consider halos composed of a minimum of 150 dark matter particles. For the
$80\:h^{-1}\,{\rm Mpc}$ simulations this corresponds to a minimum halo mass
of $5.1 \times 10^9\Msunh$ and a circular velocity $\sim 23~\kms$.

We have also carried out re-simulations of a void region in the
$50\:h^{-1}\,{\rm Mpc}$ box. This leads to a substantially improved mass
resolution with the same level of refinement. Here we achieve a mass
resolution of $8.2\times 10^6 \Msunh$ for the dark matter particles,
corresponding to a minium halo mass of $1.2\times 10^9 \Msunh$. In addition,
we have evolved this region also with the full resolution available based on
the initial high-resolution particle set-up (the $2048^3$ particle grid,
corresponding to four levels of refinement). In this case, the mass resolution
is improved to $1.0\times 10^6 \Msunh$ for the dark matter particles, and the
minium halo mass reaches down to $1.6\times 10^8 \Msunh$. We give an overview
of our simulations in Table \ref{table1}, where we also list the main
simulation parameters.

\citet{power:03} gave a simple criterion for the gravitational softening
length necessary in $N$-body simulations, $\epsilon \gtrsim r_{200}/\sqrt{N_{200}}$.
Obeying this condition ensures that particles do not suffer stochastic scattering in
the periphery of the halo which exceed the mean gravitational acceleration. 
\citet{power:03} found that even a larger softening length keeps the central
density profile unaffected. However, in a hydrodynamical simulation including radiative
cooling we wish to use a softening as small as possible to obtain an optimal resolution
for the cooled gas. Our halos are in the mass range from about $10^9 \Msunh$ to
$10^{11} \Msunh$. Using 
\begin{equation}
   M_{200}
   =
   200
   \times
   \frac{4}{3}
   \pi
   r_{200}^3
   \:
   \langle \rho \rangle
    ,
   \nonumber
\end{equation}
where $\langle \rho \rangle$ is the average cosmic matter density, we find the corresponding softening
lengths, $\epsilon \sim 2.2$ to $1\:h^{-1}\,{\rm kpc}$. Since we carry out simulations
including radiative cooling we lower the softening slightly and use for all
simulations the maximum between $2\:h^{-1}\,{\rm kpc}$ comoving and
$0.8\:h^{-1}\,{\rm kpc}$ physical. For the very high-resolution run, the
parameters $1\:h^{-1}\,{\rm kpc}$ and $0.5\:h^{-1}\,{\rm kpc}$, respectively,
are used. We have imposed a minimum SPH smoothing length equal to the
gravitational softening length.

In order to analyse in more detail the effects of the UV-background on the
baryonic content of halos, we have also carried out several additional runs of
our basic simulation of the void identified in the $50\:h^{-1}\,{\rm Mpc}$
box. Here we used different choices for the star formation and feedback
parameters, and three different values of the UV-flux normalisation, spanning
four orders of magnitude. We also ran the same simulation
without thermal stellar feedback. Using this run we can demonstrate that 
thermal feedback itself has only a minor impact on the halo baryon fraction. 
For all simulations with star formation and feedback, we selected similar
parameters for the multiphase model of the ISM as used by \citet{springel:03}.
However, we here have not included kinetic feedback (wind model) from
supernova, since then it would be difficult to disentangle the effects
caused by the UV background from those caused by supernova driven winds.

The simulations were performed on parallel supercomputers, an IBM Regatta p690
(J\"ulich Supercomputer Center, Germany), a SGI Altix 3700 (CIEMAT, Spain) and
on several different Beowulf PC clusters at the AIP and IU Bremen.  The
typical CPU time for a simulation with up to 5 million particles was $\sim 9$
CPU days on an SGI ALTIX with 32 processors. The highest resolution simulation
with 44 million particles was run on an AMD Opteron Beowulf cluster and took a
little less than 2 months of CPU time using 64 processors.

\section{Simulated dwarf galaxies in voids}

\label{sec-galax}

\begin{figure}
  \includegraphics[width=0.5\textwidth,angle=-90]{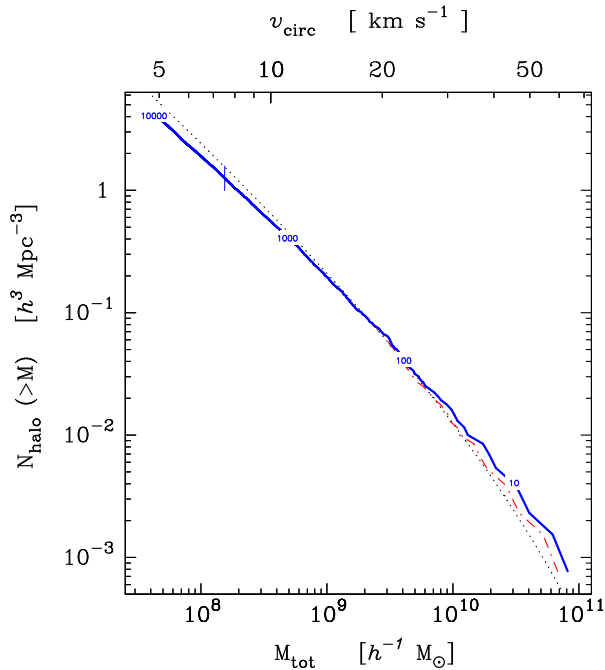}
  \caption
  { 
Mass function $n(>M)$ for our basic run halo sample at
$z=0$. The mass function is derived by taking into account the total
mass within the virial radius. The small vertical line indicates a
mass of 150 dark matter particles, which we consider as a lower limit
for an acceptable resolution. The numbers along the line indicate the
actual number of halos. The dashed line is obtained by considering the
dark matter mass instead of the total mass. For comparison the
modified Sheth-Tormen mass function derived by \citet{gottloeber:03}
is shown (dotted line). For our halo sample, the circular velocity
$v_{\rm circ} = G M_{\rm tot} / r_{\rm vir}$ can be well approximated
by $v_{\rm circ} = 31\:{\rm km \, s^{-1}} \times ( M_{\rm tot} /
10^{10}\,h^{-1}\,M_\odot )^{0.34}$.  
  }
  \label{fig-halo-mass-function}
\end{figure}

\subsection{Halo mass function}

We identify virialised halos using the Bound Density Maxima (BDM) algorithm
\citep{klypin:99}. In this method, galaxy halos are found from local density
maxima with an iterative procedure to identify the centre of mass in a small 
sphere around the centre.
Then, radial density profiles are computed. Particles that
are not gravitationally bound to the system are excluded in the computation of
the total mass. The radius of the system is selected as the minimum between
the virial radius and the point at which the density profile stops declining
(e.g.~because a nearby halo is encountered, or the halo lies within another
halo). We define the virial radius as the radius where 
the enclosed mean density equals the value
expected for a top-hat collapse model, 
\begin{equation}
   \frac{M_{\rm vir}}{4/3 \, \pi \, r_{\rm vir}^3} 
        =
   \Delta_c (z) \:
   \langle \rho \rangle
   ,
   \label{eq-def-vir-radius}
\end{equation}
where $\langle \rho \rangle$ is the mean cosmic matter density.
For the case of a flat cosmology with $\Omega_m+\Omega_\Lambda=1$,
 a useful approximation for the redshift dependent
characteristic virial overdensity is given by \citet{bryan:98}:
\begin{equation}
   \Delta_c (z)
   =
   \frac{178 + 82\,x(z) - 39 \, x^2(z) }
        { 1+ x(z)}
   ,    
   \label{eq-def-overdens}
\end{equation}
\begin{equation}
   x(z)
   =
   -
   \frac{(1-\Omega_m) \, a^3 }
        {\Omega_m  + (1-\Omega_m) \, a^3}
    , 
\end{equation}
with the cosmological expansion factor $a = 1/(z+1)$.
The mean density $\langle \rho \rangle$ evolves with redshift as
\begin{equation}
   \langle \rho \rangle (z)
   =
   \Omega_m \: \rho_{\rm crit,0} \, \frac{1}{a^3}
   =
   \Omega_m \:
   \frac{1}{a^3} \,
   \frac{3 H_0^2}
        {8 \pi G }
   .    
        \label{eq-rho-z}
\end{equation}
In low density regions, the number of interacting halos or halos with
substructure is very small. Thus, the radii of virtually all our
objects correspond to their spherical-overdensity virial radii. For
the same reason, the fraction of unbound particles in the halos is
small. Hence, we can simply consider all particles within the virial
radius to compute further halo properties.

\begin{figure}
  \begin{center}
    \includegraphics[width=0.42\textwidth,angle=-90]{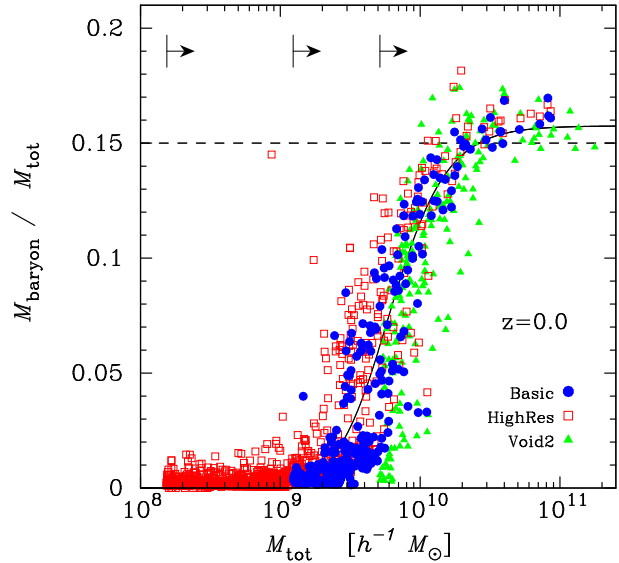}
  \end{center}    
  \caption
  { 
Baryon fraction in individual halos for differently 
resolved simulations. We compute for
each halo the baryon fraction within the virial radius,
$(M_\ast + M_{\rm gas})/(M_\ast+M_{\rm gas}+M_{\rm dm})$.
We take only those halos into account which
consist of more than 150 dark matter particles. 
From the left to the right the arrows
indicate the smallest resolved halos in the high-res, 
basic, and void2 run, respectively. 
For the high-res run we
approximate the baryon fraction by
Eq.~(\ref{eq-charact-mass}) (solid line).
}
  \label{fig-mass_baryonfraction_resolution}
\end{figure}

In Figure~\ref{fig-halo-mass-function}, we show the cumulative mass function for
our sample of void halos, based on the measured total virial masses. For
comparison, we also include a line for the dark matter halo mass function
alone, which however shows appreciable differences only for the most massive
halos. \citet{gottloeber:03} derived a modified Sheth-Tormen mass function for
halo populations in underdense regions. They showed that the mass function can
be derived from the mean density, $\Omega_{m, \rm void}$, in the volume
considered. For our halo sample analysed here (based on the high-res run), the
relevant mean density amounts to $\Omega_{m, \rm void} \simeq 0.03$. The
predicted mass function for this value is in good agreement with the one
measured for the simulated halo sample. Our simulations thus are in good agreement 
with conclusion obtained in previous works: Voids are filled with a significant
number of halos with masses $M \lesssim 10^{10} \Msunh$. If each of these
halos contains the mean cosmic baryon fraction, and cooling and star formation
within them is not suppressed significantly, a high density of luminous dwarf
galaxies should be expected in voids.

\subsection{Baryon content of dwarf galaxies}

\label{sec-baryon-frac}

In Figure~\ref{fig-mass_baryonfraction_resolution}, we show the baryonic mass fraction,
$f_b = M_{\rm baryon}(<r_{\rm vir})/ M_{\rm tot}(<r_{\rm vir})$, for each halo
identified in our simulated voids. The more massive halos in our sample,
$M_{\rm tot} \gtrsim 2\times 10^{10} \: h^{-1}\: {\rm M}_\odot$, have
approximately the cosmic mean baryon fraction, $f_{b, \rm cosm} = \Omega_b /
\Omega_m$. However, for smaller halos, the baryon fraction decreases
rapidly with decreasing halo mass.  In fact, most of the smallest halos are
nearly free of baryons.

\begin{figure}
  \begin{center}        
  \includegraphics[width=0.46\textwidth,angle=-90]{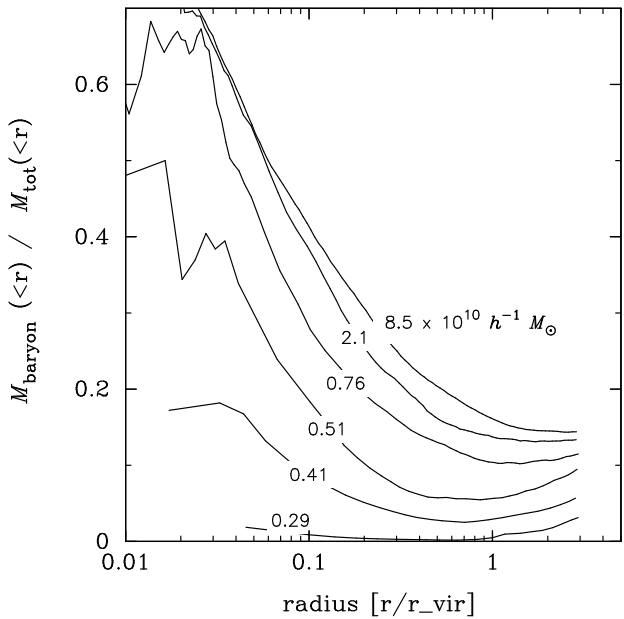}
  \end{center}
  \caption
  { 
The baryon fraction as a function of radius for halos
with different mass. All halos are chosen at $z=0$ from the basic
simulation. Radii are normalised to the BDM virial radius. Labels
along the lines indicate the mass of the halos.  
  }
  \label{fig-radius-baryonfraction}
\end{figure}

We find that the baryon fraction is insensitive to the details of the
definition of the virial radius of halos, because the cumulative baryon
fraction varies only very slowly in the outskirts of halos. This can be seen
in Figure~\ref{fig-radius-baryonfraction}, where we show the radial profile of
the baryon fraction for halos of different mass.  While the sizes of halos can
be systematically affected by different definitions of the virial radius (one
may for example choose to use only the dark matter for the definition and not
the total mass), the measured baryonic fractions are robust.

\begin{figure}
   \includegraphics[width=0.46\textwidth,angle=-90]{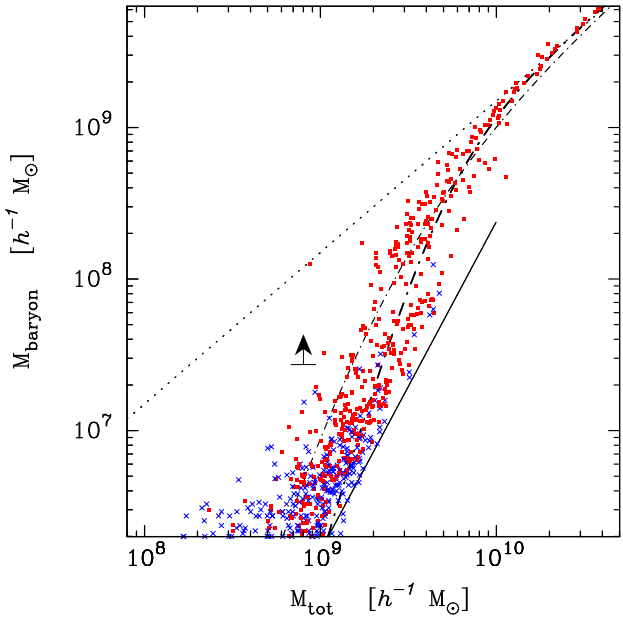}
   \caption{The baryon mass as a function of the total mass enclosed
   in the virial radius from the high-res run. Solid squares indicate
   halos with stars, and crosses those without. The dotted line shows
   the dependence if each halo would contain the mean cosmic baryon
   fraction. The solid line is derived based on the assumption that in
   low-mass halos the gas density follows the average distribution in
   the $\rho$-$T$ diagram (see Sec.~\ref{sec-empty-halos} for more
   details). The arrow indicates a baryon mass consisting of 150 SPH
   particles. The properties of smaller halos may be affected by the
   poor SPH resolution in these halos. Thin and thick dash-dotted lines show the approximation of
   Eq.~(\ref{eq-charact-mass}), with exponents $\alpha = 1$ and 2,
   respectively.  }
   \label{fig-mass-massB}
\end{figure}

We quantify the transition between the two extremes, `baryon-rich' and
`baryon-poor', using the fitting function proposed by \citet{gnedin:00},
\begin{equation}
   f_{\rm b} 
   =
   f_{b0} \:
   \left\{ 
      1+ (2^{\alpha/3} - 1 )\, \left( \frac{M_c}{M_{\rm tot}} \right)^\alpha 
   \right\}^{-3/\alpha}    
   ,
   \label{eq-charact-mass}
\end{equation}
where $f_{b0}$ is the asymptotic baryon fraction in massive halos.
Figure~\ref{fig-mass-massB} indicates that a baryon fraction decreasing with
mass can be reasonably well approximated assuming $\alpha =2$ in
Eq.~(\ref{eq-charact-mass}).  \citet{gnedin:00} found a less steep
  transition, which may be caused by radiative transfer effects included in
  his code. A partial self-shielding in halos may then reduce the radiative
  heating.  For very small halos the approximation seems to fail in any case.
  We discuss a possible origin for this in Sec.~\ref{sec-empty-halos}.

In the very massive halos of our samples we found a roughly constant baryon
fraction $f_{b0}$, independent of redshift and numerical resolution. The value
of $f_{b0}\sim 0.16$ we measured lies slightly above the cosmic mean. At the
characteristic mass, $M_c$, the baryon fraction is $f_{\rm b} = f_{b0}/2$ by
definition. For $z=0$, we derive a characteristic mass of $M_{\rm c} =
6.5\times 10^{9} \: h^{-1}\:{\rm M}_\odot$ from the `high-res' run.

  It is important to consider whether numerical resolution effects
  influence this result. In particular, a too small number of resolution
  elements per halo could easily introduce a spurious baryon reduction due to
  numerical oversmoothing. The resulting characteristic mass would then depend
  on the number of particles in a halo rather than on the halo mass itself
  when simulations with quite different mass resolutions are compared. In
  Fig.~\ref{fig-mass_baryonfraction_resolution}, we compare the void2, basic
  and high-res runs. Their baryon fractions as a function of halo mass overlap
  nicely, even though the mass resolution differs by more than an order of
  magnitude.  Therefore, significant numerical oversmoothing occurs only in
  halos smaller than those included in our analysis.

  The marginal displacement of the characteristic mass as a function of
  resolution can be understood in terms of the better resolved merging
  histories of higher resolution runs. The better the mass resolution, the
  more small progenitors of a final halo can be resolved and contribute stars
  to the final object. In particular, many more star particles are formed
  while the UV-background radiation is still vanishingly small. The resulting
  difference can be seen in Tab.~1: The basic and the high-res run have
  identical initial conditions, except for the increased resolution, which
  leads to the production of $\sim 10\%$ more stars.  As a result, the
  characteristic mass of the high-res run is also slightly lowered.  However,
  this effect is limited by the fraction of mass in progenitors which can form
  stars before reionisation.

\begin{figure}
  \begin{center}
    \includegraphics[width=0.45\textwidth,angle=0]{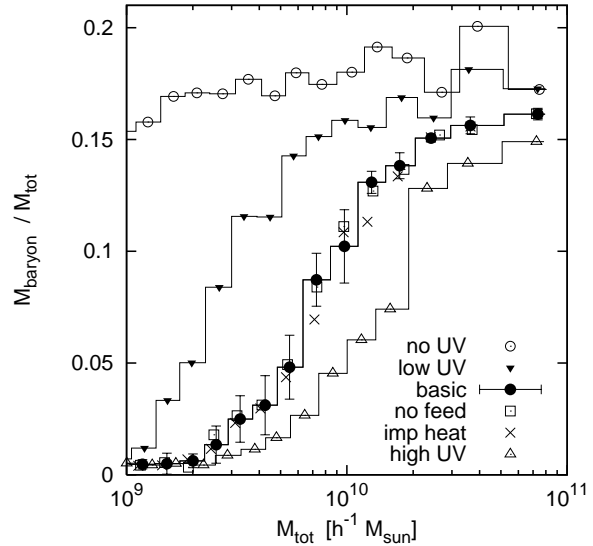}
  \end{center}    
  \caption
  {  
Baryon fraction as a function of mass for different UV-fluxes.
The photo-heating by the UV-background is varied from
zero to 
$J_0^{\rm UV} = 95 \times 10^{-23}\:{\rm ergs \, 
s^{-1} \, cm^{-2} \, sr^{-1} \, Hz^{-1}}$,
see Tab.~1.
The average baryon fraction in mass bins is computed; for
the basic-run also the r.m.s. deviation is shown by error bars.
In addition baryon fractions are depicted for the impulsive heat
model with $1.5\times 10^4 \: {\rm K}$ (imp heat) and for the 
simulation without thermal stellar feedback (no-feed).
  }     
  \label{fig-mass_baryonfraction_UV}
\end{figure}

  The importance of the UV-background for the baryon fraction is
  demonstrated in Fig.~\ref{fig-mass_baryonfraction_UV}. We have carried out
  three simulations for which we multiplied our standard UV-flux for every
  redshift by the factors 0.0, 0.01, and 100, keeping the spectral shape of
  the UV-background radiation constant.  These strong variations of the
  UV-background flux level displace the characteristic mass-scale only by a
  factor $\lesssim 3$. 
  We will discuss in the next section, Sec~\ref{sec-how-to-supp}, 
  that this shift is caused by 
  the modification of the equilibrium temperature.
  The run with zero UV-flux clearly shows that
  the background radiation causes the baryon deficit: In this case, even the
  smallest halos show the average cosmic baryon fraction.

\begin{figure*}
  \begin{center}
  \includegraphics[width=0.3\textwidth,angle=-90]{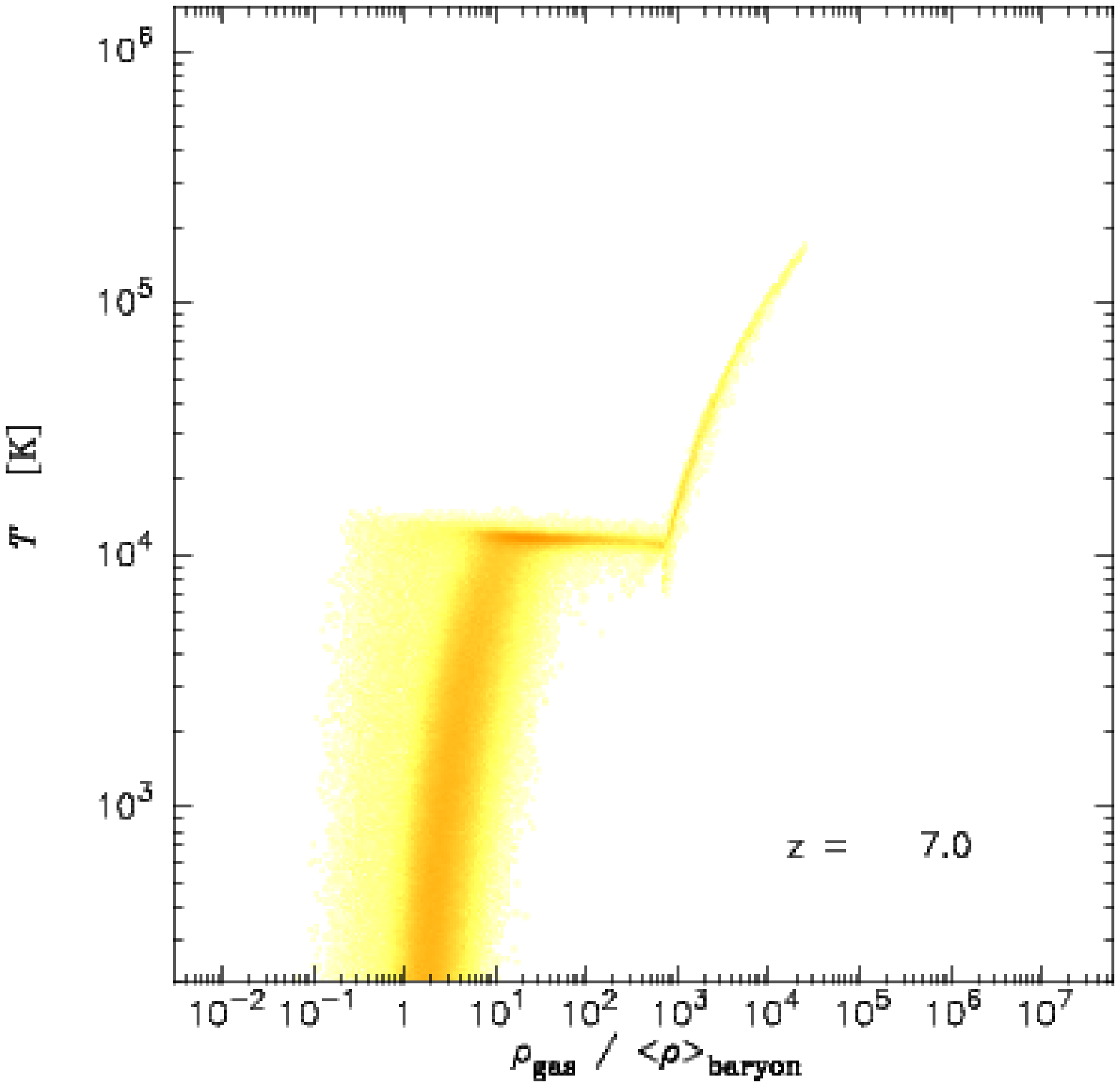}
       \hfill
  \includegraphics[width=0.3\textwidth,angle=-90]{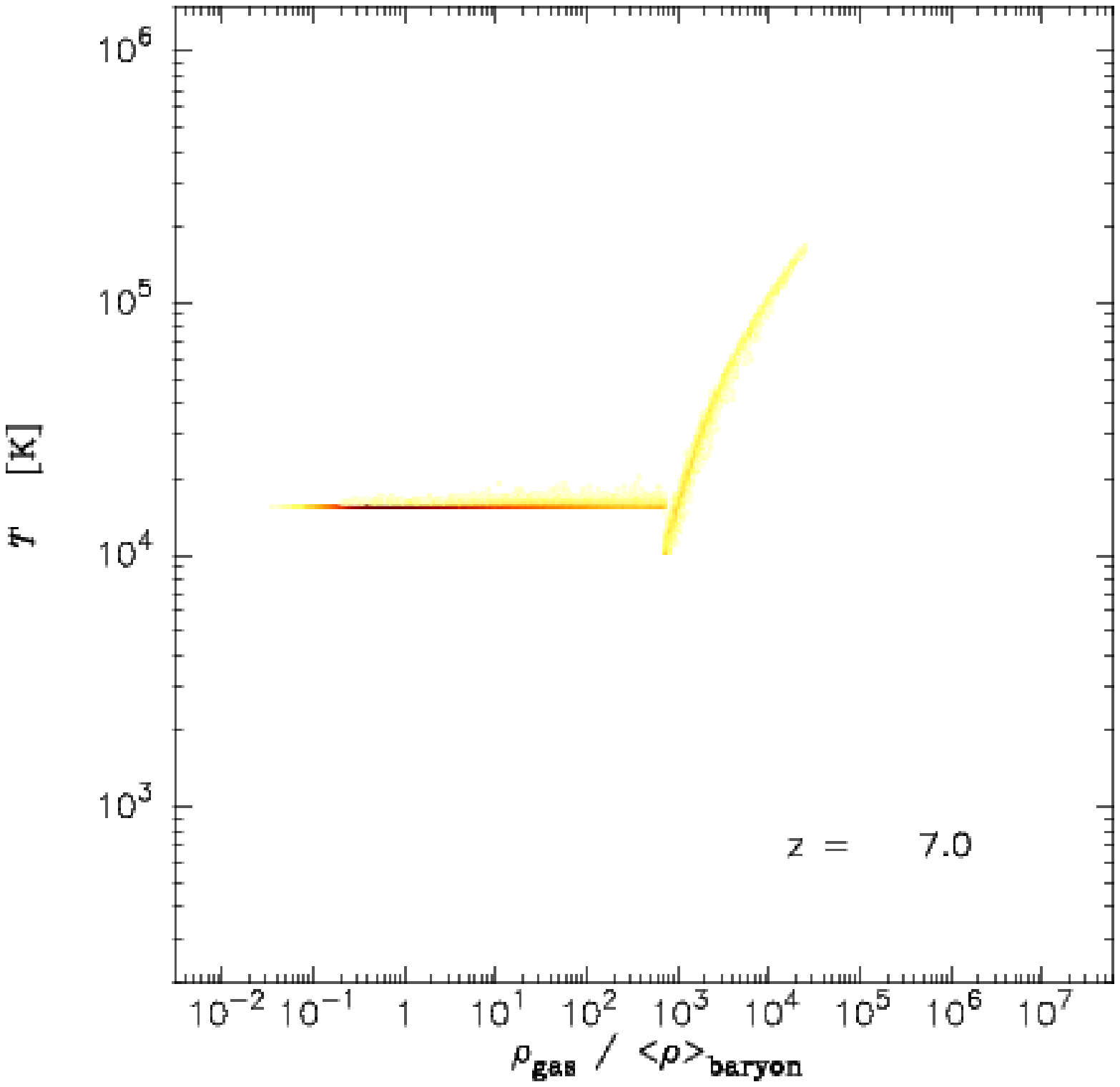}
       \hfill
  \includegraphics[width=0.3\textwidth,angle=-90]{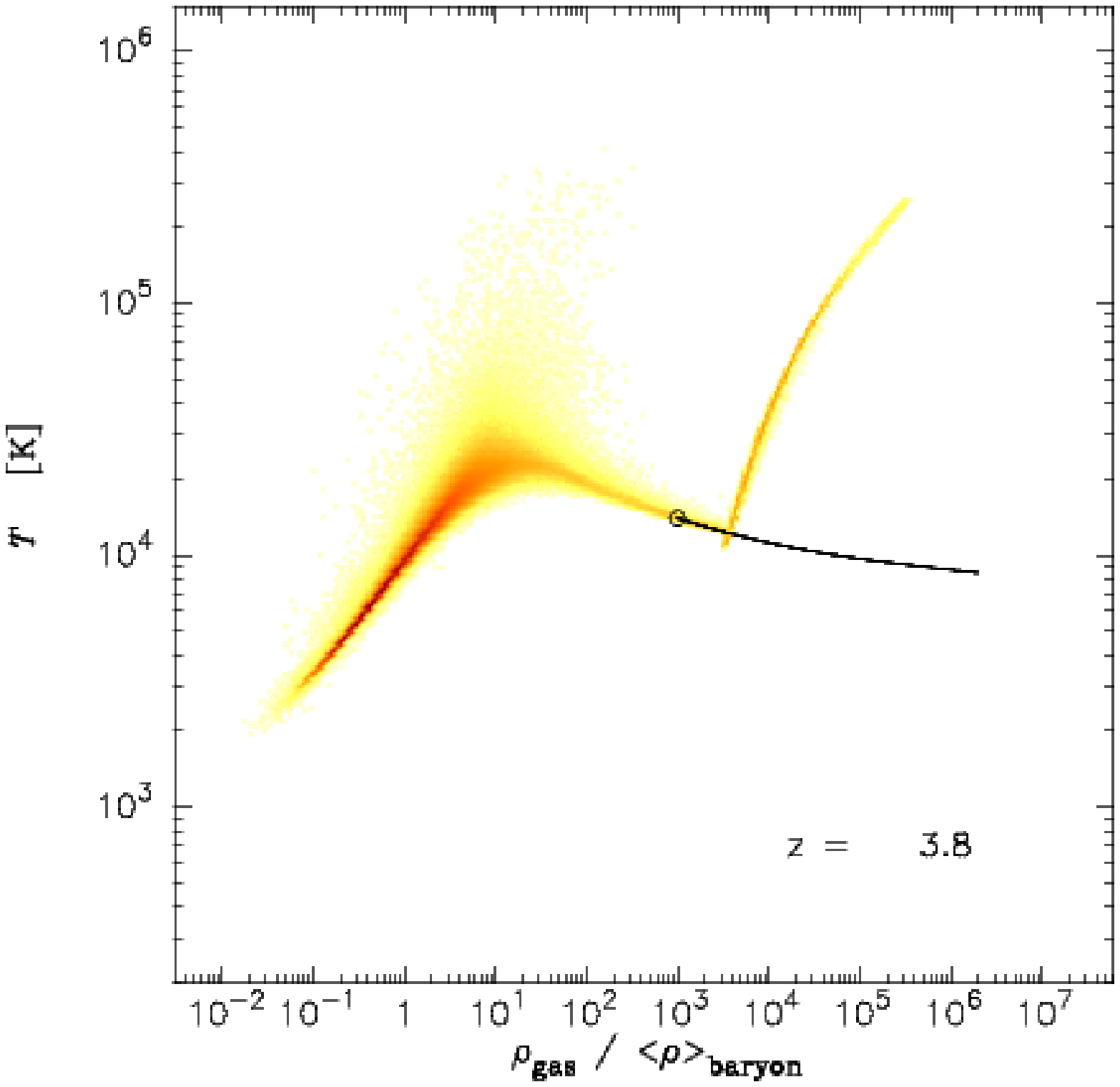}
  \end{center}
  \caption{
The distribution of particles in the density -
temperature phase diagram for different redshifts for the
impulse-heat-15 simulation.
From $z=7$ to 6 the minimum gas temperature is set to $1.5\times10^4 \:{\rm K}$.
For smaller redshift the standard heating-cooling scheme is applied.
At $z\sim4$ the initial heat pulse is faded away by adiabatic cooling
for densities $\rho_{\rm gas} / \langle \rho \rangle_{\rm baryon} \gtrsim 0.1$. 
Equilibrium and entry temperatures are indicated by the solid lines and open
circles, respectively. See Sec.~\ref{sec-how-to-supp} and Fig.~\ref{fig-J-Teq}
for a discussion of them.
 }
  \label{fig-rho_T_phase_diagram_impulse}
\end{figure*}

  During the epoch of reionisation, the true heat input may be larger than
  computed in our heating-cooling scheme, which is based on the assumption of
  collisional ionisation equilibrium. In the onset of reionisation,
  non-equilibrium effects can be significant, however, and the ionised fraction
  increases very rapidly. While our implicit solver for the evolution of the
  thermal energy deals gracefully with this situation, part of the injected
  energy may be missed if the ionized fraction jumps from zero to a finite
  value in the course of one time step. One may speculate that the heat pulse
  at reionisation is underestimated by our method, but that it may have a
  significant effect on the baryon fraction in the halos later on if fully
  taken into account. 
  In order to test this question, we have carried out two 
  simulations which mimic an upper limit for the impulsive heat input: from
  $z=7$ to 6, all gas is heated to a minium temperature of $1.5\times
  10^4\:{\rm K}$ and to $5.0\times10^4\:{\rm K}$ in the runs imp-heat-15 and
  imp-heat-50, respectively. After z=6 the gas evolves again according to our 
  standard heating-cooling scheme, see Fig.~\ref{fig-rho_T_phase_diagram_impulse}.
  Even with a strong heat input at the epoch of reionisation, our measured
  characteristic mass at $z=0$ is hardly affected, as can be seen in
  Fig.~\ref{fig-mass_baryonfraction_UV}. Moreover, the different heat input at
  reionisation has virtually no effect on the evolution of the 
  characteristic mass. This is caused by the short cooling times at high redshift:
  Fully ionised gas with average cosmic density and a temperature of $5.0\times10^4\:{\rm K}$
  has at $z=6$ a cooling time as short as $\sim 50\:{\rm Myr}$, because the maximum of 
  hydrogen line cooling is in this temperature range. Inverse Compton cooling with CMB photons
  is also very efficient at these redshifts. It keeps the cooling times short even if we would heat
  to a temperature above the efficient line cooling.  
  Since gas in the halo of a galaxy 
  is much denser, the cooling time is even shorter. Hence, the energy injected by a heat-pulse
  at the epoch of reionisation is radiated away on time-scales smaller than the dynamical ones.

Finally, efficient stellar feedback can in principle also remove gas from small
halos.  However, for the thermal feedback considered in the multiphase
feedback model used in our simulations, such a gas removal does not
occur. While the feedback regulates the consumption of cold gas by
star formation, it does not cause gaseous outflows. The latter only
occur in our simulations if explicitly modeled with a kinetic feedback
component \citep{springel:03}.  However, we deliberately avoided the
inclusion of such feedback models in this study, allowing us to focus
on the impact of UV-heating in a clean fashion.
Figure~\ref{fig-mass_baryonfraction_UV}  verifies that in
simulations without feedback the characteristic mass scale is at the same place as in
the basic run.

\begin{figure}
   \includegraphics[width=0.46\textwidth,angle=-90]{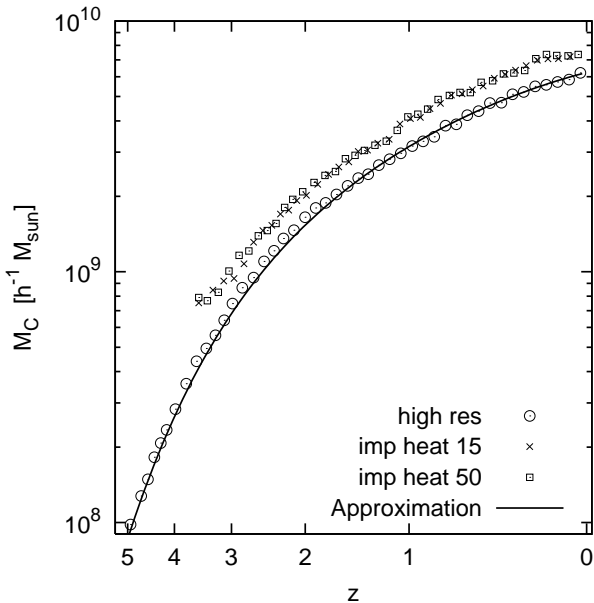}
   \caption{Evolution of the characteristic mass scale $M_{\rm c}$ for
   several simulations. We approximate the halo baryon fractions for
   each simulation by Eq.~(\ref{eq-charact-mass}). We determine
   $M_{\rm c}$ using a least-squares method. $f_{b0}$ is a free
   parameter, but it is almost constant for all simulations, $f_{b0}
   \sim 0.16$. Open circles indicate the results
   for the high-res simulations. Here the smallest halos included 
   have $1.5 \times 10^8 \: h^{-1} \: M_\odot$. Crosses and squares show
   the results for imp-heat-15 and imp-heat-50 simulations, respectively.
   Here the smallest halos included have $10^9 \: h^{-1} \: M_\odot$. Note 
   that the characteristic mass may lie somewhat below the mass of the smallest halos,
   since it is derived from a fit to the baryon fractions.
   The solid line shows the approximation for $M_{\rm c}(z)$ from the
   high-res run using Eq.~(\ref{eq-charact-mass-evolve}) and
   (\ref{eq-def-tau}).  }
   \label{fig-massF}
\end{figure}

In order to determine the evolution of characteristic mass with redshift, we use a
least-square fit of Eq.~(\ref{eq-charact-mass}) to our measurements of the
baryon mass fraction at a number of different simulation output times.  We can
infer the transition mass scale reasonably well from our high-res simulation up to
$z\sim5$. In Fig.~\ref{fig-massF}, we show the resulting evolution of
$M_c(z)$. Interestingly, as structure grows towards lower redshifts,
progressively more massive halos become baryon-poor.

  It is interesting to also compare this result  with the time
  evolution of $M_c$ measured for the simulation with an additional heat pulse
  at the epoch of reionisation, which is also shown in Figure~\ref{fig-massF}.
  The two curves are identical except for a small offset. As discussed above,
  the latter can be attributed to the different mass resolutions of the two
  simulations.  The similarity of the two curves therefore shows that the heat
  pulse at reionisation does at most weakly influence the characteristic mass
  at $z\lesssim3$.  $M_c(z)$ is relatively insensitive at these redshifts to
  the previous thermal history of the low density gas at high redshift, but it
  depends more strongly on the current state of the gas and that of the UV
  background.

The evolution of the characteristic mass $M_{\rm c}(z)$ can be expressed as
\begin{equation}
   \frac{ M_{\rm c} (z) }{ 10^{10} \: h^{-1} \: M_\odot }
   =
        \left\{ \tau(z) \:
                \frac{ 1} {1+z}
        \right\}^{3/2} \:
        \left\{
                \frac{\Delta_c(0)}{\Delta_c(z)}
        \right\} ^ {1/2}        
   ,
   \label{eq-charact-mass-evolve}
\end{equation}
were $\tau(z)$ encodes the evolution of the minimum virial temperature
required for halos to still allow further cooling in the presence of the 
UV-background.  We will discuss this criterion and its derivation in more detail
below. Treating $\tau(z)$ as a simple analytic fitting function for the
moment, we find that our numerical results can be well described by
\begin{equation}
        \tau(z) 
        = 
        0.73 \times
        (z+1)^{0.18} \,
        \exp \{ - ( 0.25\, z )^{2.1} \}
  \label{eq-def-tau}
\end{equation}
for the `high-res' run over the entire redshift range, see
Fig.~\ref{fig-massF}.

\subsection{How to suppress gas condensation}

\label{sec-how-to-supp}

A quantitative understanding of the characteristic mass can be obtained by
considering the equilibrium line between photo-heating rate and cooling rate
in the density-temperature phase-space plane.  In
Figure~\ref{fig-rho_T_phase_diagram}, we show the gas particles at three
different times, and include this equilibrium line for high-density gas,
computed self-consistently from the cooling and heating routines in \G.  Much
of the gas in the density range $\delta \sim  10^3 - 10^6$ is indeed distributed 
along this line.

We define the equilibrium temperature at a fiducial overdensity of 1000 as
``entry temperature'' $T_{\rm entry}$ into the condensed phase. Gas that
condenses in a halo will at least reach this temperature due to photo-heating
before it can cool further to the $\simeq 10^4\,{\rm K}$ reached at very high
overdensities. Note that this is independent of the potential difference
between a ``cold mode'' or a ``hot mode'' of accretion \citep{keres:04}.  In
the cold mode, gas creeps along the lowest possible temperature into the
condensed phase, without being heated by an accretion shock to the virial
temperature first, as it happens in the ``hot mode''. However, even in the
cold mode, the gas will at least be heated to the ``entry temperature''
$T_{\rm entry}$ by the UV-background.  After reaching this temperature, it can
then evolve along the equilibrium line towards higher densities and eventually
reach the onset of star formation. In a sense, for halos with $T_{\rm vir} =
T_{\rm entry}$, the hot and cold mode should therefore become largely
identical.

We argue that a comparison of $T_{\rm vir}$ with $T_{\rm entry}$ provides
a simple criterion that tells us whether the gas in a halo can still cool. To
demonstrate this, we show that this assumption provides a quantitative
explanation for our measurements of $M_c(z)$.
We define the virial temperature for our halos as
\begin{equation}
        k_{\rm B} T_{\rm vir}
        =
        \frac{1}{2}
        \mu
        m_{\rm p}
        \frac
                {G      M_{\rm vir} }
                {r_{\rm vir}}
        ,               
        \label{eq-T-vir}
\end{equation}          
where $m_{\rm p}$ is the proton mass and $k_{\rm B}$ is the Boltzmann constant. 
The mean molecular weight of the fully
ionised gas is $\mu=0.59$. Virial mass and radius are related by the definitions
of Eqs.(\ref{eq-def-vir-radius}) and
(\ref{eq-rho-z}). Hence the virial temperature depends on the mass as
\begin{equation}
        T_{\rm vir}
        =
        \frac  {1}  {2}
        \frac  {\mu m_p}  {k_B}
        \left\{
                \frac  {\Delta_c (z) \Omega_m}  {2}
        \right\}^{\frac{1}{3}}          
        (1+z)
        \left\{
                G
                M_{\rm vir}
           H_0  
        \right\}^{\frac{2}{3}}
        .
        \label{eq-Tvir-def}
\end{equation}
For the cosmological concordance model used here we insert $\Omega_m=0.3$ and
cast Eq.~(\ref{eq-Tvir-def}) into 
\begin{equation}
        T_{\rm vir}
        =
        3.5 \times 10^{4}\:
        {\rm K} \;
        (1+z)
        \left\{
                \frac  { \Delta_c (z) }  { \Delta_c (0) }
        \right\}^{\frac{1}{3}}
        \left\{
                \frac  { M_{\rm vir} }  { 10^{10} h^{-1} M_\odot }
        \right\}^{\frac{2}{3}}
        .
        \label{eq-Tvir-values}
\end{equation}
Now we apply our criterion introduced above: If the virial temperature of a halo is
below the entry temperature into the condensed phase, its
potential well is not deep enough to compress the gas sufficiently and to
overcome the pressure barrier generated by the photo-heating. Thus, we expect
that the characteristic halo mass necessary to increase the condensed baryonic
mass can be estimated by rewriting Eq.~(\ref{eq-Tvir-values})  into
\begin{equation}
   \frac{M_{\rm c}(z)}{ 10^{10} \: h^{-1} M_\odot}
   \simeq
   \left\{
      \frac  { T_{\rm entry}(z) }  { 3.5 \times 10^{4} \: {\rm K} }
      \frac  { 1 }   { 1+z }
   \right\}^{\frac{3}{2}}
   \left\{
      \frac  { \Delta_c (0) }  { \Delta_c (z) }
   \right\}^{\frac{1}{2}}
   .
   \label{eq-Mc-theo}
\end{equation}
  This relation can only be an approximation, since $M_c(z)$ will in
  general depend on the entire history of a halo, whereas $T_{\rm entry}$ is
  an instantaneous quantity. However, we have seen from the `imp-heat'
  simulation that the characteristic mass is mainly determined by the current
  state of the gas. We therefore assume that $M_c(z)$ can be considered as an
  instantaneous quantity. In this sense, Eq.~(\ref{eq-Mc-theo}) gives the
  motivation for our ansatz of Eq.(\ref{eq-charact-mass-evolve}). Also it
  shows that in a first approximation the fitting function $\tau(z)$ is
  proportional to $T_{\rm entry}(z)$.

\begin{figure}
  \begin{center}
  \includegraphics[width=0.46\textwidth,angle=-90]{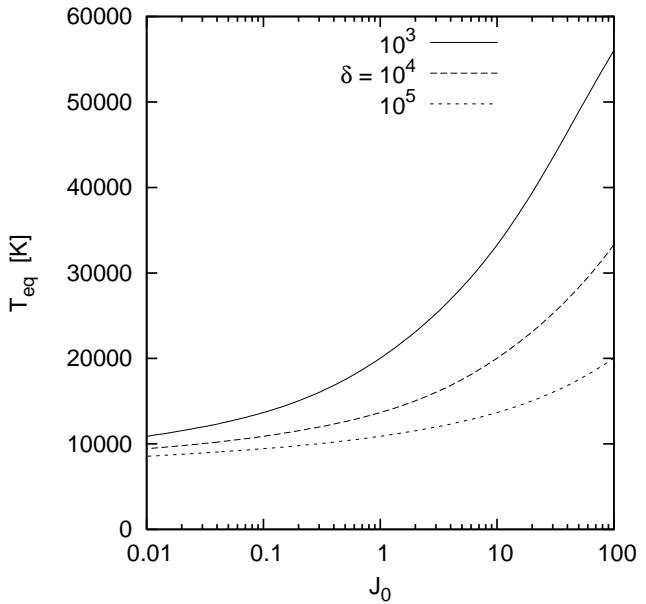}
  \end{center}
  \caption
  { 
The equilibrium temperature, $T_{\rm eq}$, as a function of the amplitude, $J_0$, of the UV-background
radiation.  $T_{\rm eq}$ is computed by solving ${\cal H}(T_{\rm eq},\rho) = \Gamma(T_{\rm eq},\rho)$.
Heating and cooling functions as implemented for the numerical simulations are used. The amplitude dependence
is given for three densities, $\rho = \delta \Omega_{\rm b} \rho_{\rm crit}$, for $z=0$.
  }
  \label{fig-J-Teq}
\end{figure}

In Figures~\ref{fig-rho_T_phase_diagram} and \ref{fig-rho_T_phase_diagram_impulse} solid lines
indicate the equilibrium temperature, ${\cal H}(T_{\rm eq},\rho) = \Gamma(T_{\rm eq},\rho)$,  
for $\delta > 10^3$. In that density region cooling times 
are short and the gas is essentially distributed along the equilibrium line --or according to 
the multi-phase ISM prescription used in {\sc Gadget-2}. The equilibrium temperature depends on the UV-background.
The dependency on the amplitude of the background radiation, $J_0$, is shown in Fig.~\ref{fig-J-Teq}.
If we lower or increase the UV-flux by two orders of
magnitude, the entry temperature shifts roughly by a factor of two. With $M_c \propto T_{\rm entry}^{3/2}$ 
we obtain the shift of $M_c$ as found in the simulations, see Sec.~\ref{sec-baryon-frac} 
and Fig.~\ref{fig-mass_baryonfraction_UV}.

\begin{figure}
  \begin{center}
  \includegraphics[width=0.46\textwidth,angle=-90]{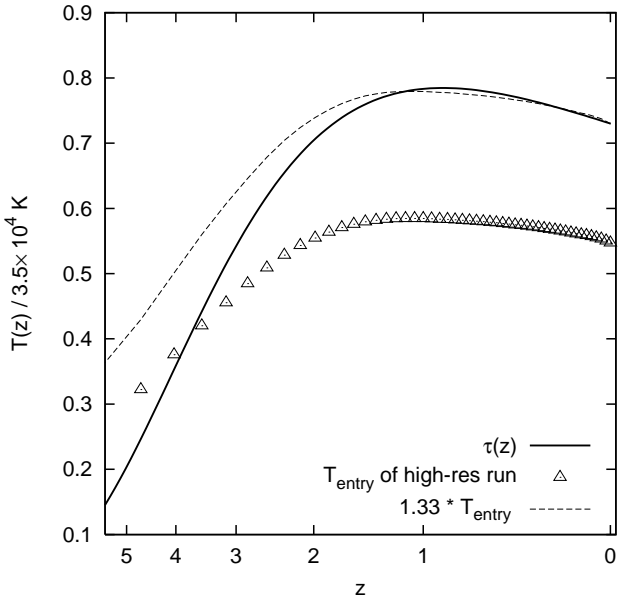}
  \end{center}
  \caption
  { 
Evolution of the entry temperature. We derive from
the high-res run the entry temperature, $T_{\rm entry}$, for several
redshifts (open triangles). $T_{\rm entry}$ is determined according to
the description in the caption of
Fig.~\ref{fig-rho_T_phase_diagram}. For comparison we plot the
expression $\tau(z)$ given in Eq.~(\ref{eq-def-tau}) (solid line). The
scaled entry temperature (``refined model'') is also shown (dashed
line).  
  }
  \label{fig-temp-entry}
\end{figure}

In the following, we use the abbreviation $\tilde{T}_{\rm entry} = T_{\rm
  entry}(z) / 3.5\times 10^4\:{\rm K}$. In Figure~\ref{fig-temp-entry}, we
show the evolution of $\tilde{T}_{\rm entry}(z)$ for the high-res run and
compare it with the expression for $\tau(z)$ derived from fitting the measured
redshift evolution of the characteristic mass in the same simulation. First of
all, we note that the redshift variation of both is small compared with the
evolution of the characteristic mass. This indicates that the large variation
of the latter is mainly governed by the $1/(z+1)$ term in
Eq.~(\ref{eq-charact-mass-evolve}).  Furthermore, from
Fig.~\ref{fig-temp-entry} we can conclude that the simple criterion invoked
above reproduces the evolution of $M_{\rm c}(z)$ up to $z\sim5$ astonishingly
well, with a deviation of a few $10\:\%$.

However, the condition $T_{\rm vir} = T_{\rm entry}$ appears to underestimate
the characteristic mass for small redshift to some extent. This is not
surprising for a number of reasons. For example, we compute the virial radius
from the dark matter distribution only, hence the total masses used here are
slightly above the virial masses. Furthermore, $T_{\rm entry}(z)$ determined
at a fixed overdensity 1000 may underestimate the true entry temperature.  We
can account for this and improve our match of the numerical results by
refining our criterion in the following way: The virial temperature of a halo
has to be $\gtrsim 1.3 \times T_{\rm entry}$ to permit further condensation.

To explain the deviation at high redshifts we have to acknowledge the fact
discussed above: $\tau(z)$ and $\tilde{T}_{\rm entry}$ are completely different in the
sense that the latter denotes which halos are able to add more gas to the
condensed phase at a given time, while the former reflects the entire
evolution history of dwarf halos. Before reionisation, no halo is
photo-evaporated, i.e. $M_c = \tau = 0$. It takes some time after reionisation 
to produce baryon-poor halos. Hence, $\tau(z)$ should be below $T_{\rm entry}$ 
for some time after reionisation.

In any case, the agreement demonstrated in Figure~\ref{fig-temp-entry} shows that
the argument that the virial temperature should at least equal the entry
temperature provides a good quantitative description for the transition scale
between baryon-poor and baryon-rich halos.  Interestingly, this explanation
can account for the effect simply by alluding to the ongoing photo-heating of
the gas by the UV-background radiation, and the accompanying increase of the
gas pressure in low mass halos. At $z\lesssim 3$ the effects of
the impulsive heating during the epoch of reionisation play only a
subdominant role.

\subsection{Baryon deficit in the linear theory}

Small-scale baryonic fluctuations grow slower than the corresponding dark
matter fluctuations due to the counteracting pressure gradients. To describe
this effect,
\citet{gnedin:98} introduced a filtering wavenumber $k_{\rm F}$ over which
baryonic fluctuations are smoothed out.  \citet{gnedin:00} compared the
corresponding filtering mass
\begin{equation}
M_{\rm F}
        =
        \frac{4 \pi}{3}
        \langle \rho \rangle
        \left( \frac{ 2 \pi a }{k_{\rm F}}
        \right)^3
        \label{eq-mF}
\end{equation}
up to $z=4$ with the characteristic mass measured in 
cosmological reionisation simulations and found a good agreement. 

\begin{figure}
  \includegraphics[width=0.4\textwidth,angle=-90]{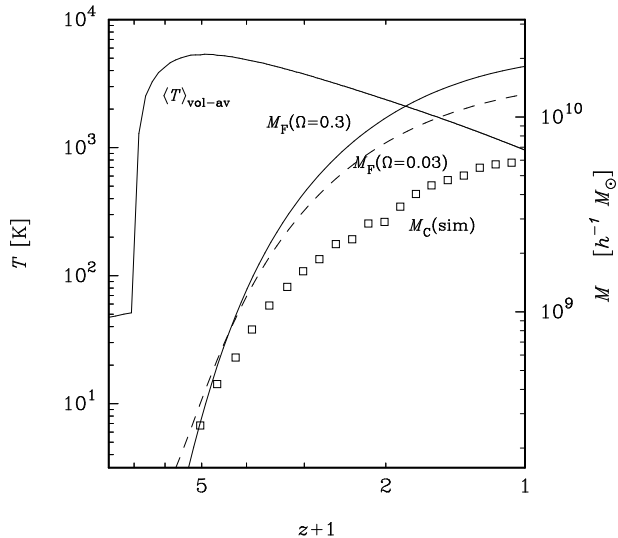}
  \caption 
  {
Comparison of filtering and characteristic mass. The thin line indicates the evolution of the 
volume averaged temperature in the simulation. The thick solid line gives the
filtering mass integrated according to Eq.~(\ref{eq-k-filter-a}) using the
averaged temperature depicted here. The dashed line shows the filtering mass
for a low density region with $\Omega_m = 0.03$. Open squares indicate the
characteristic mass obtained from the high-res simulation.
  }
  \label{fig-massF-lin}
\end{figure}

\citet{gnedin:98} showed that $k_{\rm F}$ can be related to the linear growth
function $D(t)$ by
\begin{equation}
\frac{1}{k_{\rm F}^2}
        =
        \frac{1}{D}
        \int\nolimits_0^t dt' \:
        \frac{c_{\rm s}^2 ( \ddot{D} + 2 H \dot{D} )}{ 4 \pi G \langle \rho \rangle } \:
        \int\nolimits_{t'}^t dt'' \:
        \frac{1}{a^2}
        ,
        \label{eq-k-filter}
\end{equation}
with the time-dependent Hubble parameter $H = \dot{a}/a$. We rewrite
Eq.~(\ref{eq-k-filter}) in order to integrate it up to $z=0$.
The linear growth function obeys the relation 
\begin{equation}
\ddot{D}
        +
        2 H \dot{D}
        = 
        4 \pi G  \langle \rho \rangle \: D
        ,
        \label{eq-D}
\end{equation}
which we can use to remove the time derivatives in Eq.~(\ref{eq-k-filter}). 
We also substitute the integration variable by $a$. With 
\begin{equation}
\left(\frac{H(a)}{H_0}\right)^2 
        = 
        S^2(a) 
        = 
        1 
        + 
        \Omega_m \left(
                \frac{1}{a}-1
                \right) 
        +
        \Omega_\Lambda(a^2+1)
        \label{eq-H}
\end{equation}
we then get
\begin{equation}
\frac{1}{k_{\rm F}^2}
        =
        \frac{1}{D}
        \int\nolimits_0^a da' \:
        \frac{c_{\rm s}^2}{H_0^2} \:
        \frac{D}{S} \:
        \int\nolimits_{a'}^{a} da''
        \frac{1}{a''^2 S}
        \label{eq-k-filter-a}
\end{equation}
For a given cosmology the linear growth function can be computed by \citep{carroll:92}
\begin{equation}
D(a)
        =
        \frac{5}{2}
        \Omega_m \:
        \frac{S(a)}{a} \:
        \int\nolimits_0^a da' \:
        \frac{1}{S^3(a)}
        ,
        \label{eq-D-a}
\end{equation}
hence we obtain a solution for Eq.~(\ref{eq-k-filter}) for any redshift,
provided the evolution of the sound speed, $c_{\rm s}(a)$, is known.  To obtain
the sound speed for the high-resolution region in the simulations we follow
again \citet{gnedin:98} and compute the volume averaged temperature,
\begin{equation}
c_{\rm s}(a)
        =
        \frac{5}{3}
        \frac{k_{\rm B} \langle T \rangle_\mathrm{vol}(a)}{\mu m_{\rm p }}
        .
        \label{eq-cs}
\end{equation}
Figure~\ref{fig-massF-lin} shows that $\langle T \rangle_{\rm vol}$ rises
sharply at the time of reionisation, thereafter it decreases slowly from $\sim
5000 \: {\rm K}$ to $1000 \: {\rm K}$. These temperatures are lower than
expected for an average cosmic volume because the considered void region is
underdense and the temperature scales with density (see
Fig.~\ref{fig-rho_T_phase_diagram}).

In Fig.~\ref{fig-massF-lin}, we show the resulting filtering mass scale.  For
redshift $z=0$, it is about three times larger than the characteristic mass
obtained from the simulations, while the difference becomes progressively
smaller towards higher redshift. One may argue that the underdensity of the
void region has to be taken into account. To consider a rather extreme case we
assume that the void evolves like an open universe with $\Omega_m = 0.03$ and
unchanged $\Omega_\Lambda=0.7$. As shown in Fig.~\ref{fig-massF-lin} (dashed
line) the density of the region has only a moderate effect on the filtering
mass. It reduces to some extent the difference with our measured characteristic masses.
At a first glance one may expect a different density dependence of $M_{\rm F}$.  
The average density enters Eq.~(\ref{eq-mF}) directly via its appearance in the 
denominator of Eq.~(\ref{eq-k-filter}). If the density did not affect the Hubble 
constant, $H$, and the linear growth function, $D$, we would expect the filtering mass is
proportional to $\langle \rho \rangle^{-1/2}$. However, $H$ and $D$ depend on $\langle \rho \rangle$
as can be seen from Eq.~(\ref{eq-D}). In order to derive consistently the filtering mass
as a function of  $\Omega_m$ the density dependence of $H$ and $D$, as given in Eqs.~(\ref{eq-H}) 
and (\ref{eq-D-a}), has to be taken into account. 
For high redshift, $a \ll 1$, we can approximate
\[
 	S^2(a) 
	\sim
	\Omega_m
	\frac{1}{a}
	.
\]
This leads to 
\[
        \frac{1}{k_{\rm F}^2}
        \propto
        \frac{1}{\Omega_m}
\]
and hence
\[
	M_{\rm F}
	\propto
	\Omega_m^{-1/2}
	.
\]
For high redshift we obtain the same proportionality as expected without taking the
 contribution of $H$ and $D$ into account. However, the result changes for lower redshifts. 
For $a=1$ we have $S=1$, i.e. there is no dependence on $\Omega_m$. This implies that the integral Eq.~(\ref{eq-k-filter-a}) depends less strong on $\Omega_m$ and in $M_{\rm F}$ the factor $\langle \rho \rangle$ becomes dominant.

In summary, we expect that at high redshift the filtering mass shows the proportionality $M_{\rm F} \propto \Omega_m^{-1/2}$, whereas for low redshift even the opposite behavior may occur. The actual dependence can only be found by computing the integral in Eq.~(\ref{eq-k-filter}).
Note that the speed of sound $c_s$ is also a function of redshift and modifies the integral additionally. The dashed line in Fig.~\ref{fig-massF-lin} gives the result for a low density region with $\Omega_m = 0.03$ and for the rather low average temperature in the void region. At low redshift the filtering mass overpredicts significantly the characteristic mass we measure.

The density dependence in Eq. (\ref{eq-mF}) is partially
compensated by the fact that for $a\ll1$ the filtering wave number scales with
density according to $k_{\rm F}^2 \propto \Omega_m$.  
In conclusion, we find
that at low redshift the filtering mass $M_{\rm F}$ overpredicts the
characteristic masses we measure.

\subsection{Are there dark halos free of baryons?}

\label{sec-empty-halos}

We argued above that photo-heating may prevent further gas condensation in
dwarf halos. If it never takes place in a given halo, or an existing condensed
phase has been evaporated at some time by the UV-background, the halo will
only contain diffuse gas, where ``diffuse'' for our purposes denotes gas of
too low density to support star formation. It is therefore also interesting to
distinguish between halos with and without stars, as we have done in
Fig.~\ref{fig-mass-massB}. This allows us to demonstrate again how closely the
equilibrium line in the $\rho$-$T$ diagram, the virial temperature, and the
baryon fraction are connected.

The expected baryonic mass in a halo is given by
\begin{equation}
        M_{\rm b}
        =
        \frac{4}{3} \:
        \pi \:
        r_{\rm vir}^3 \:
        \overline{\rho}_{\rm b}
        ,
        \label{eq-mass-b}
\end{equation}
where $\overline{\rho}_{\rm b}$ denotes the average gas density in a halo. At
redshift $z=0$, the majority of the gas is distributed at low densities along the
power-law relation
\begin{equation}
        \left( 
                        \frac{\rho_{\rm eq}}
                                {\langle\rho\rangle_{\rm b}}
        \right)^{0.57}
        =
        \frac{T_{\rm eq}}{3.6 \times 10^3 \:{\rm K}}
        \label{eq-low-dens-pow-law}
\end{equation}  
in the $\rho$-$T$ phase diagram, where photo-heating balances adiabatic cooling
due to the expansion of the universe. If we assume that the average
temperature in halos is given by the virial temperature, then the average
density (of the diffuse gas) in a halo cannot lie below the relation given by
Eq.~(\ref{eq-low-dens-pow-law}), i.e.~this imposes a rough upper limit on the
average gas density, and hence the diffuse baryonic content of the halo. With
the help of Eq.~(\ref{eq-Tvir-values}), we can replace the virial temperature
by the virial mass and cast this condition for the average gas density in a
halo into
\begin{equation}
        \left( 
                        \frac{\overline{\rho}_{\rm b}}
                                {\langle\rho\rangle_{\rm b}}
        \right)^{0.57}
        =
        9.7 \times
        (z+1)
        \left\{
                \frac  { \Delta_c (z) }  { \Delta_c (0) }
        \right\}^{\frac{1}{3}}
        \left\{
                \frac  { M_{\rm vir} }  { 10^{10} h^{-1} M_\odot }
        \right\}^{\frac{2}{3}}
        \label{eq-mean-dens-in-halo}
        .
\end{equation}  
Plugging this result into Eq.~(\ref{eq-mass-b}) and using the definition of the
virial radius, Eq.~(\ref{eq-def-vir-radius}), we obtain 
\begin{equation}
        \frac  { M_{\rm b} }  
          { 10^{10} h^{-1} M_\odot }
        =
        \frac{84.7}{\Delta_c(0)} \:
        \frac{\Omega_b}{\Omega_m} \:
        \left\{
                \frac  { M_{\rm vir} }  { 10^{10} h^{-1} M_\odot }
        \right\}^{2.17}
        .
        \label{eq-massB-in-halo}
\end{equation}
for the baryon mass at $z=0$.
The resulting baryon mass is shown in Fig.~\ref{fig-mass-massB}. Halos without
any stellar matter are primarily distributed just above this line. Hence
Eq.~(\ref{eq-massB-in-halo}) provides a rough upper limit for the diffuse
baryonic mass in dwarf-sized halos. In summary, for halos which contain
virtually no condensed baryons, the diffuse gas mass can be determined from
the equilibrium line in the $\rho$-$T$ phase diagram. This can hence explain
why halos significantly below $M_{\rm c}$ are seemingly almost empty of gas.
It provides also a lower limit for the baryon fraction.

\begin{figure*}
   \begin{center}
   \includegraphics[width=0.52\textwidth,angle=-90]{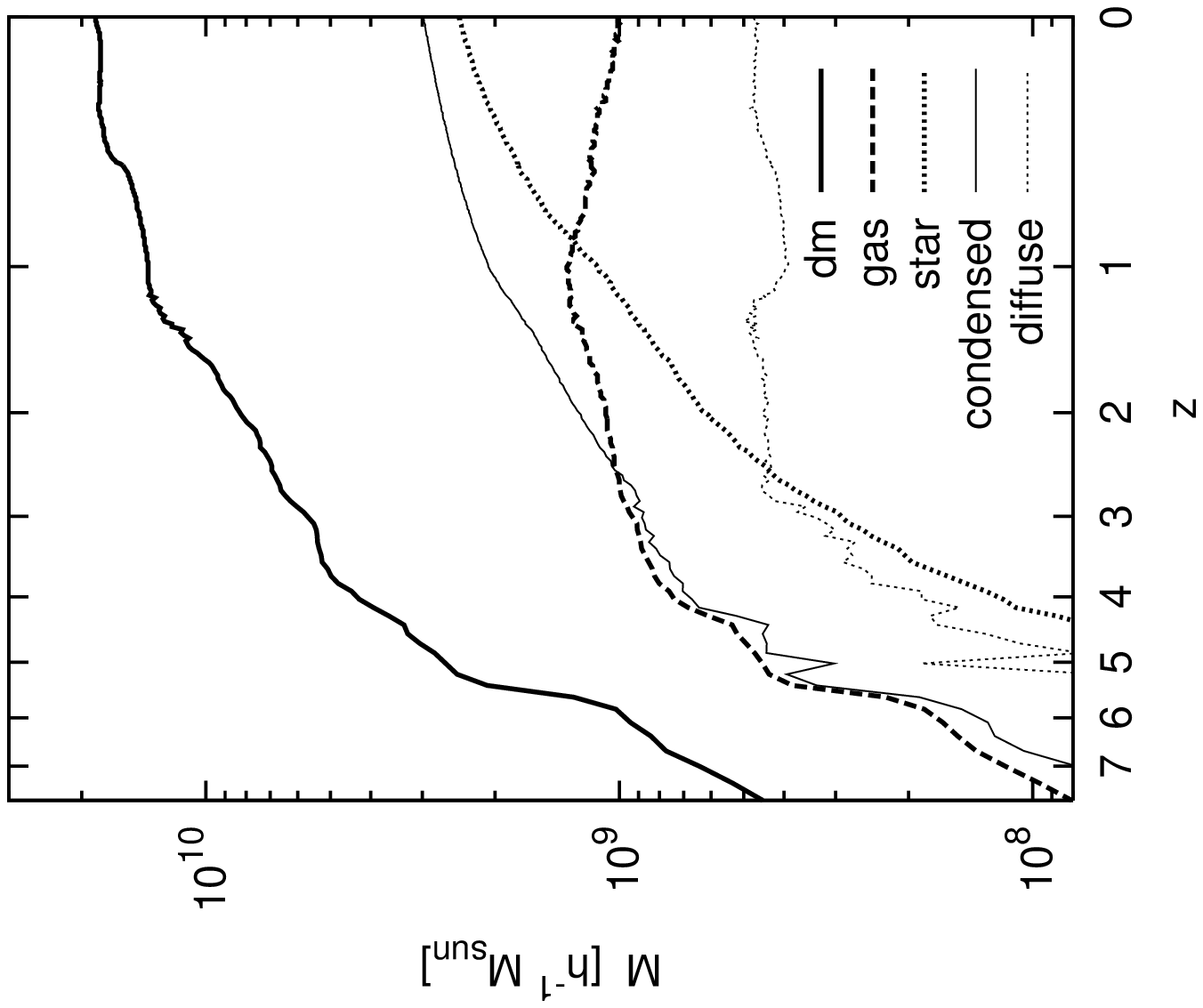}
   \includegraphics[width=0.52\textwidth,angle=-90]{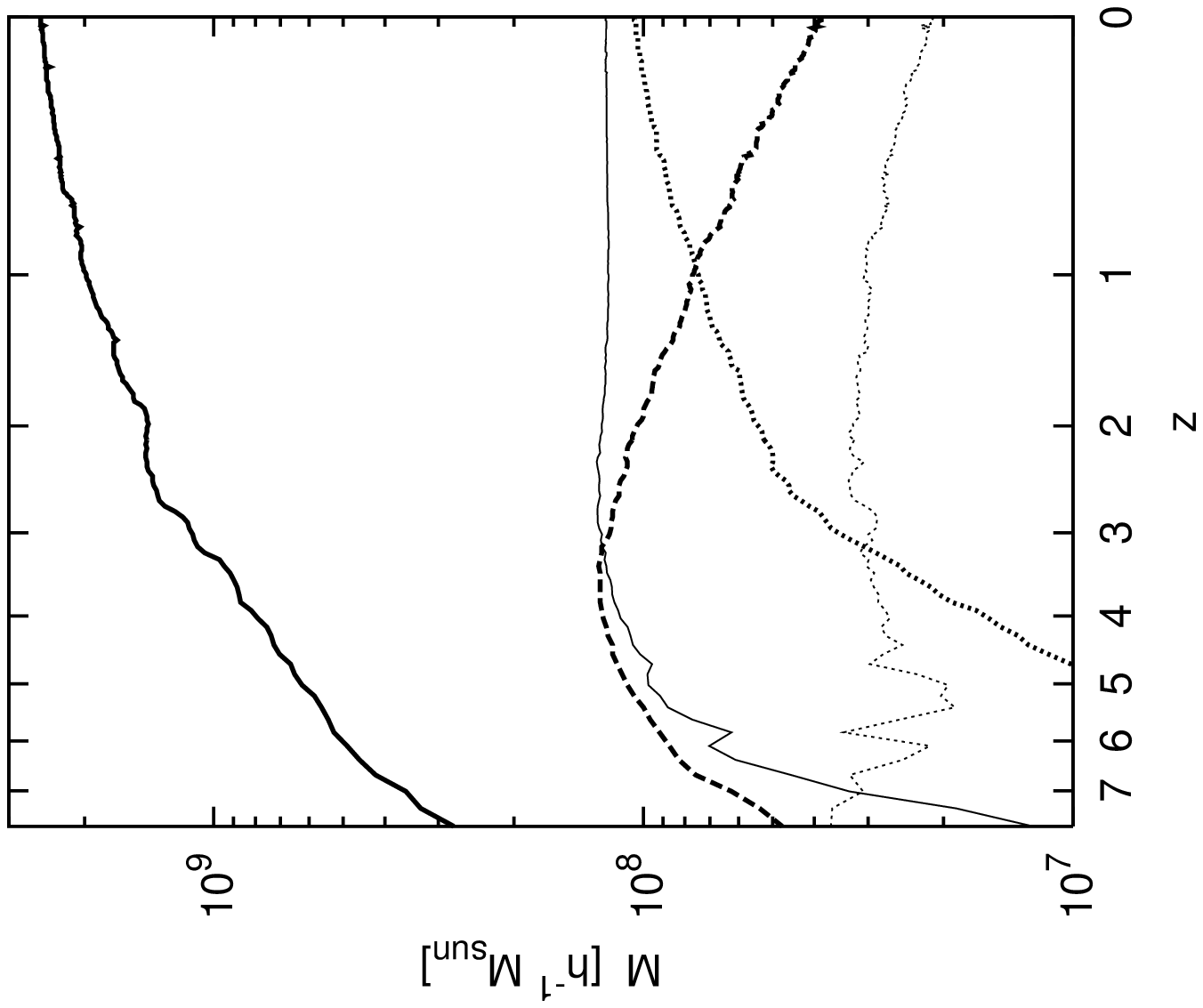}
   \end{center}
   \caption{Evolution of the mass in different components for two
high-resolution halos, one above $M_{\rm c}$ (left panel) and one
below $M_{\rm c}$ (right panel). The evolution of the total mass
(thick solid line), the condensed gas mass (thick dashed), the stellar
mass (thick dotted) and diffuse gas mass (thin dotted) is shown. In
addition the evolution of the condensed mass (gas+stars) is shown
(thin solid line).  }
   \label{fig-mass-accretion-exempl}
\end{figure*}

\subsection{Condensation history}

The basic argument above is that photo-ionisation can stop at some
time the condensation process of gas in dwarf galaxy halos. We can
support this idea further by finding halos with a constant amount of
mass in the condensed phase.  Figure~\ref{fig-mass-accretion-exempl}
shows the evolution of the different mass components for two example
halos. A halo selected with mass significantly above $M_{\rm c}$ at
$z=0$ has continuously increased the dark matter, stellar, and
condensed masses, while the gas mass decreased.  In a halo with total
mass significantly below $M_{\rm c}$, the individual mass components
change as well, but in contrast to the case of the more massive halo,
the condensed mass is remarkably constant.  This lends strong support
to our basic argument. We see here that for the individual halo shown
neither further gas condensates nor photo-heating evaporates all the
gas.  Thus even in the optically thin treatment of the galactic gas
reservoir, photo-heating is not able to expel gas from the galaxy that
has overdensities above $\sim 1000$.  Instead, the condensed gas
continues to be slowly converted into stars. It is thus not expected
that photo-heating instantaneously switches off star formation.

\begin{figure*}
   \begin{center}
   \includegraphics[width=0.52\textwidth,angle=-90]{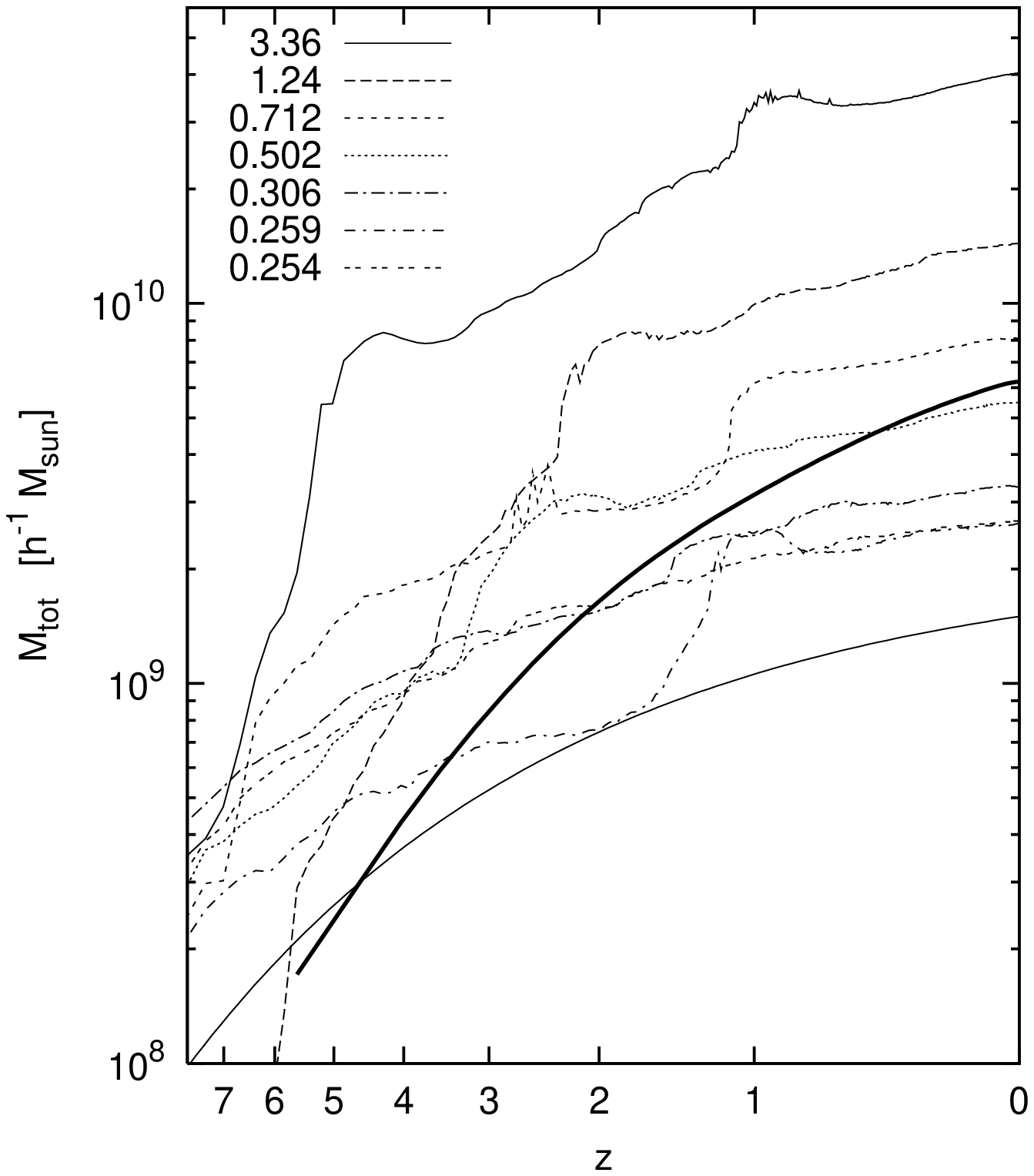}
        \hspace{-2cm}
        \hfill
   \includegraphics[width=0.52\textwidth,angle=-90]{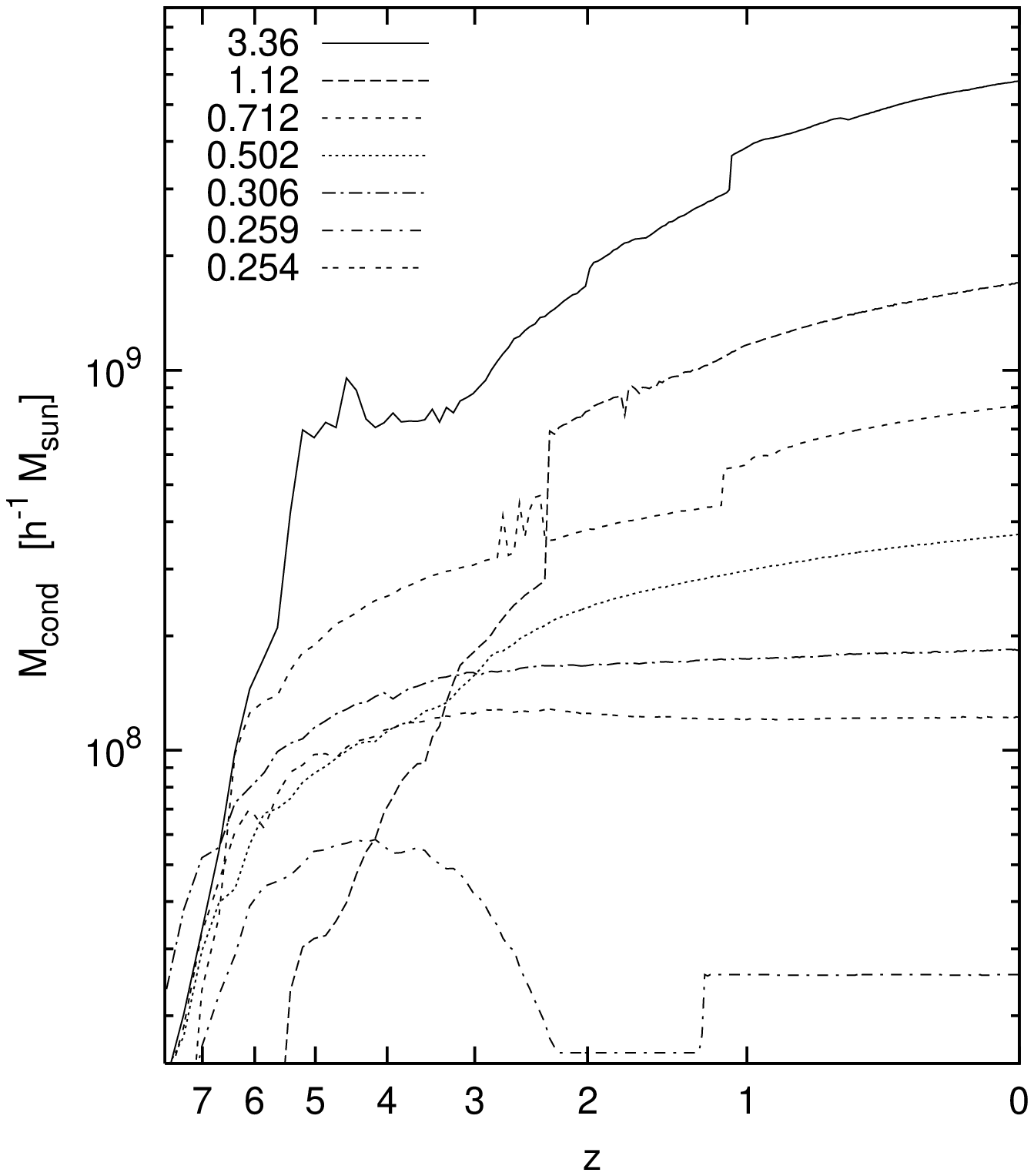}
   \end{center}
   \caption{ Mass accretion histories. We compute the mass accretion
histories by choosing from the high-res run several halos at $z=0$ and
searching repeatedly for the most massive progenitor. The left panel
shows the evolution of the total mass, while the right panel shows the
evolution of the condensed baryon mass of the same halos. Jumps in the
condensed mass indicate merger events. The numbers indicate the dark
matter mass at $z=0$ in $10^{10}\:h^{-1}\,M_\odot$. In the left panel,
the evolution of the characteristic mass for the `high-res' run is
also shown (thick solid line). For comparison, we also plot the mean
mass accretion history (thin solid line) for $M_{\rm tot} = 1.4 \times
10^9 \: h^{-1} \, M_\odot$ with $\alpha = 0.35$, as expected for a low
density environment, see Eq.~(\ref{eq-MAH}).  }
   \label{fig-mass-accretion-hist}
\end{figure*}

We now analyse the mass accretion histories of several halos of different mass.
In Fig.~\ref{fig-mass-accretion-hist}, we show that the evolution of the
condensed phase changes systematically from halo masses below $M_{\rm c}$ to
those above. For the latter, the total mass and the condensed mass grow almost
in parallel. For halos with mass close to $M_{\rm c}$, the condensed mass
increases monotonically but slower than the total mass. Finally, for halos
with mass significantly below $M_{\rm c}$, the condensed mass remains
constant. Thus the evolution of individual halos is perfectly consistent with
the result that photo-heating primarily stops condensation in small halos.
Moreover, we find that our ``refined'' criterion for condensation reproduces
the time at which condensation stops quite well. To show this explicitly we
have plotted in the left panel of Fig.~\ref{fig-mass-accretion-hist} $M_{\rm
  c}$ derived form $T_{\rm vir} = 1.3 \times T_{\rm entry}$. The times at
which the low-mass halos fall below $M_{\rm c}$ correspond to the times at
which the condensed masses begin to remain constant (right panel).

In halos with $M_{\rm tot}(0) \lesssim 2\times
10^{9} h^{-1} M_\odot$ the condensed gas phase is apparently evaporated completely. 
Even if these halo appear to be sufficiently
well resolved at $z=0$, we caution that condensation can only take
place at $z\lesssim 4$, probably at even higher redshift since $M(z) <
M_c(z)$.  Hence we expect that in a simulation with higher resolution
the condensed phase might stay and result in a noticeably larger
amount of stars.  However, we cannot really exclude that condensed gas
in halos with $M_{\rm tot}(0) \lesssim 2\times 10^{9} h^{-1} M_\odot$
becomes generally unstable at some time and is therefore evaporated
even for arbitrary good numerical resolution.  Such an assumption is
frequently invoked in simple analytic treatments. For instance, the
analysis of \citet{barkana:99} is based on the assumption that all gas
with $T>T_{\rm vir}$ is evaporated.  The fact that the significantly
improved resolution of the `high-res' run compared to the `basic' run
does not lead to a stable condensed gas phase in all halos with mass
about $M_{\rm tot}(0) \sim 2\times 10^{9} h^{-1} M_\odot$ seems to
point into this direction.  However, more stringent test of this will
require better resolution for the progenitor of these halos.

\subsection{Stellar mass function}

Using our estimates for the suppression of baryon condensation in halos with
mass below $M_{\rm c}(z)$, an approximation for the expected stellar content
in small halos may be derived. To this end we assume a mean mass accretion
history for the small halos in our sample. \citet{wechsler:02} and
\citet{bosch:02} showed 
that the average dark matter mass accretion histories can be
approximated by
\begin{equation}
        M_{\rm MAH}(z)
        =
        M_{\rm tot}(z=0) \:
        e^{-\alpha z }
        ,
        \label{eq-MAH}
\end{equation}
which works especially well for halos with an early formation time. In
Fig.~\ref{fig-mass-accretion-hist}, we show an average accretion history
(thin solid line) using $\alpha=0.35$. 
Moreover, from the work of \citet{bosch:02} one can derive an appropriate
value for $\alpha$ for halos with $\sim 5\times 10^{9}\:h^{-1}\,M_\odot$ in a
low density universe.  

All halos in our sample have rather flat accretion histories compared
to the evolution of the characteristic mass $M_{\rm c}(z)$, thus most
of them have been able to condensate baryons and to form stars at high
redshift. But at some time the halo mass fell below $M_{\rm c}(z)$,
and from this time onwards the mass in the condensed phase is expected
to remain constant.  To test this, we estimate the stellar mass in a
halo by computing first the redshift $z^{\rm eq}$ at which the mass
according to the accretion history equals the characteristic mass,
$M_{\rm MAH}(z^{\rm eq}) = M_{\rm c}(z^{\rm eq})$. At this redshift,
we determine the condensed mass, assuming here that at this time the
baryon fraction in the halos has the cosmic mean value
$\Omega_b/\Omega_m$. In addition, we assume that on average $80\:\%$
of the baryons are in the condensed phase, see Fig.~\ref{fig-mass-accretion-exempl}. 
Finally, as an upper limit for the amount of
stars that may form, we assume that all of the condensed baryons are
eventually converted into stars, $\langle f_{\rm eq} \rangle = 0.8$.
Hence we estimate the stellar mass in a given halo as $M_{\rm eq}(z=0)
= \langle f_{\rm eq} \rangle\: \Omega_b/\Omega_m \: M_{\rm MAH}
(z^{\rm eq})$.  For halos more massive than $M_c(z=0)$ the maximum
stellar mass is given just by $\langle f_{\rm eq} \rangle\:
\Omega_b/\Omega_m \: M_{\rm tot}(0)$.

\begin{figure}
   \begin{center}
   \includegraphics[width=0.45\textwidth,angle=-90]{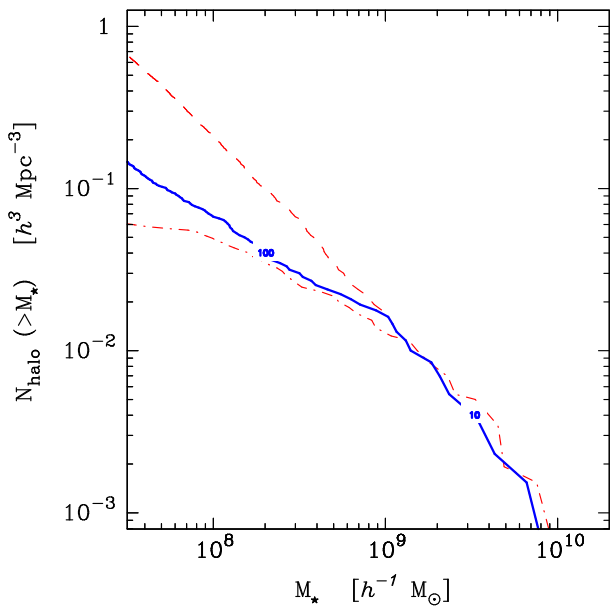}
   \end{center}
   \caption{ Stellar mass function at $z=0$. We assign a stellar mass
to each halo in the following manner: if $M_{\rm tot}(z=0) > M_c(0)$
the stellar mass is $\langle f_\ast \rangle \Omega_b / \Omega_m \:
M_{\rm tot}$. For smaller halos we compute the redshift $z^\ast$ at
which $M_{\rm tot} \exp( - \alpha z^\ast ) = M_c(z^\ast)$. We compute
the stellar mass according to $\langle f_\ast \rangle \Omega_b /
\Omega_m \: M(z^\ast)$. The solid line show the results for the halo
sample from the basic run and the parameters $\langle f_\ast \rangle =
0.8$ and $\alpha = 0.35$. Numbers along the line indicate how many
halos are actually in our void sample. For comparison the stellar mass
function obtained from the simulation (dash-dotted) is shown. In
addition, we show the result for the assumption that in each halo the
stellar mass amounts to $\langle f_\ast \rangle \Omega_b / \Omega_m \:
M_{\rm tot}$ (dashed line).  }
   \label{fig-star-mass-function}
\end{figure}

We define the stellar mass function, $N_{\rm halo}(M_\ast)$, as the
number of halos with stellar mass larger than $M_\ast$, independent of
the dark matter masses of the halos that host the galaxies.
Figure~\ref{fig-star-mass-function} shows the stellar mass function
obtained by the procedure described above (solid line). For
comparison, we also show the mass function for a fixed stellar mass
fraction $f_\ast$ for all halos (dashed line). The deviation of the
two curves below $M_\ast \sim 10^9\:h^{-1}\:{\rm M}_\odot$ indicates
that in low-mass halos the stellar mass fraction falls below the fixed
value $f_\ast$, hence those halos are shifted to smaller ${M}_\ast$
and the mass function begins to flatten. For a stellar mass content of
$\sim 10^8\:h^{-1}\:{\rm M}_\odot$ the number of halos is reduced by a
factor of three.  The simulated halo sample (dash-dotted line)
provides a lower limit for the stellar mass function because of the
limited numerical resolution.  Above $\sim 10^8\:h^{-1}\:{\rm
M}_\odot$, the two curves agree very well. For smaller masses, the two
curves deviate since some halos have formed too few stars due to
insufficient resolution. In conclusion, the photo-evaporation reduces
the number of halos with $M_\ast \gtrsim 10^8\:h^{-1}\:{\rm M}_\odot$
by a factor of about four and, more importantly, the faint-end slope
of the stellar mass function is much shallower than expected based on
the dark halo mass function of $N$-body simulations alone.

\section{Discussion}

\label{sec-discuss}

The main purpose of this work has been to investigate the effect of UV-heating
on the baryon fraction and the stellar mass of dwarf-sized isolated halos in
voids. This is motivated by the suggestion that photo-heating may provide a
potential explanation for the observed low number density of galaxies in
voids, and the apparent paucity of luminous galactic satellites in the Local Group, 
in contrast to the large abundance of dark matter halos and subhalos, respectively, 
predicted by CDM models.

In order to examine this question, we have carried out some of the presently
best resolved hydrodynamical simulations of galaxy formation in large regions
of space. Our simulations are unique in the sense that they specifically
sample cosmological void regions at high resolution and evolve them to the
present epoch.

We have identified a characteristic mass scale below which the baryon fraction
in a halo is reduced by photo-heating. This happens for masses $M_{\rm
  vir}(z=0) \lesssim 6.5\times 10^{9}\:h^{-1}\:M_\odot$ (denoted here as dwarf
galaxies), and this characteristic mass scale depends only weakly on the
UV-flux. The suppression itself happens due to photo-evaporation of gas out of
halos, thereby reducing the baryon fraction of dwarf galaxies, and also by
offsetting cooling losses in halos such that further condensation of baryons
is strongly reduced. We have derived a simple quantitative criterion that
gives the characteristic mass $M_{\rm c}$ at which the baryon fraction is
on average half that of the cosmic mean.  This is phrased on a condition on
the virial temperature which must at least reach an ``entry temperature'' in the
instable branch of the cooling/heating equilibrium line in the gas phase-space
diagram.  Our prediction for $M_{\rm c}(z)$ is strikingly well in accord with
the evolution obtained from the simulations.

At redshift $z=0$, the characteristic mass scale of photo-evaporation
corresponds to a circular velocity of $\sim 27 \: {\rm km \, s^{-1}}$,
which is significantly less than we would obtain with the filtering
mass formalism introduced by \citet{gnedin:00} using an entirely
different approach. We thus predict a considerably milder effect of
photo-heating on the evolution of small halos. This has significant
consequences for semi-analytic models of galaxy formation that relied
on photo-heating as a feedback mechanism to suppress small
galaxies. For instance, \citet{somerville:02} concluded that the
substructure problem in the Local Group can be easily solved taking
photo-evaporation into account, provided the effect is as strong as
suggested by the analysis of \citet{gnedin:00}. This conclusion is
probably no longer valid when the effect is much weaker, as we find
here.

It is clear however that the baryon fraction and hence the star fraction rate
in void dwarf galaxies is significantly reduced by the UV-background. As a
result, the halo mass function cannot be translated directly into a mass
function of the luminous matter. In the framework of the physics included in
our simulations, we argue that the baryonic mass belonging to a halo is
essentially set at the time when $M_{\rm vir}(z) = M_{\rm c}(z)$. We compute
the resulting mass function of the stellar content in halos under this
assumption and obtain good agreement with our direct simulation measurements.

We caution however that it is numerically difficult to resolve all the star
formation in small dwarf galaxies at high redshift.  While we have found some
dwarf halos in our simulation without any stars, it seems plausible that this
is due to insufficient resolution, because all halos with a progenitor with
$M(z) > M_{\rm c}(z)$ at some $z$ should have been able to create some stars.
Only in those halos with a very steep mass accretion history, steep enough to
be for all redshifts below $M_{\rm c}(z)$, star formation should always be
suppressed.

Our results are derived under the assumption of optically thin gas with a
spatially uniform UV flux, neglecting any effects due to self-shielding in the
inner parts of halos. Recently, \cite{susa:04} analysed photo-evaporation with
3D radiation transfer simulations for individual halos. They found that for
both a full radiation transfer treatment and a optical thin approximation the
fast rising UV-background at high redshift efficiently suppresses star
formation. However, a full radiation transfer treatment with self-shielding
can only increase the amount of stars because it makes the UV heating less
efficient. Neglecting self-shielding in our analysis may thus lead to an {\em
  overestimate} of the effects of the UV-background.

The evolution of the characteristic mass function depends essentially on the UV-flux history. We have used here the \citet{haardt:96} model, where reionisation takes place at redshift $z=6$.  Previous  results from WMAP \citep{spergel:03} suggested  that the Universe could have been reionised at much higher redshift. If true, the evolution of $M_{\rm c}$ could be much more shallow at high redshift. Consequently, more halos would have an accretion history entirely below $M_{\rm c}(z)$, and hence more halos would remain virtually free of condensed baryons. The most recent analysis from three year data of  WMAP \citep{spergel:06} indicates that reionisation took place in the redshift rage $z\sim 11$ to $7$. This would lead to a moderately steeper $M_{\rm c}(z)$. Interestingly, the so modified characteristic mass function would evolve almost parallel to the average accretion history, allowing both, halos which condensate gas only at high redshifts and others which start late to condensate. However, our measurements of the characteristic mass function based on  the standard reionisation  at $z=6$ seem to be  a good approximation  for the effects of UV-heating  in  evaporating baryons  of dwarf size halos.

In summary, it appears unlikely that adjustments in the reionisation
history can alter our basic finding that UV-heating is not particularly
efficient in evaporating all baryons out of dwarf-sized halos. This makes it
questionable whether feedback by photo-heating is really sufficient to suppress
dwarf galaxies in voids, luminous satellites in galaxies, and to flatten the
faint end of the galaxy luminosity function as much as observationally indicated. 
Other proposed solutions like
kinetic supernova feedback and associated galactic winds may therefore be
needed to resolve these problems completely.

\section*{Acknowledgements}
This work has been partially supported by the Acciones Integradas
Hispano-Alemanas.  MH and GY thanks financial support from the Spanish
{\em Plan Nacional de Astronomia y Astrofisica} under project number
AYA2003-07468. This research was supported in part by the National
Science Foundation under Grant No.  PHY99-0794. GY and SG thanks the
Kavli Institut for Theoretical Physics for hospitality. We thank the
John von Neumann Institute for Computing (Germany), the CIEMAT at
Madrid (Spain), and the CLAMV at IU~Bremen for kindly allowing us to
use their computational facilities.

\newcommand{\aap  }{A\&A}
\newcommand{\araa }{ARA\&A}
\newcommand{\apj  }{ApJ}
\newcommand{\apjs }{ApJS}
\newcommand{\apjl }{ApJL}
\newcommand{\apss }{ApSS}
\newcommand{\aapr }{A\&A~Rev.}
\newcommand{\aj   }{AJ}
\newcommand{\mnras}{MNRAS}

\bibliographystyle{mn2e}

\bibliography{void}

\begin{thebibliography}{}

\bibitem[\protect\citeauthoryear{{Babul} \& {Rees}}{{Babul} \&
  {Rees}}{1992}]{babul:92}
{Babul} A.,  {Rees} M.~J.,  1992, \mnras, 255, 346

\bibitem[\protect\citeauthoryear{{Barkana} \& {Loeb}}{{Barkana} \&
  {Loeb}}{1999}]{barkana:99}
{Barkana} R.,  {Loeb} A.,  1999, \apj, 523, 54

\bibitem[\protect\citeauthoryear{{Bryan} \& {Norman}}{{Bryan} \&
  {Norman}}{1998}]{bryan:98}
{Bryan} G.~L.,  {Norman} M.~L.,  1998, \apj, 495, 80

\bibitem[\protect\citeauthoryear{{Carroll}, {Press} \& {Turner}}{{Carroll}
  et~al.}{1992}]{carroll:92}
{Carroll} S.~M.,  {Press} W.~H.,    {Turner} E.~L.,  1992, \araa, 30, 499

\bibitem[\protect\citeauthoryear{{Couchman} \& {Rees}}{{Couchman} \&
  {Rees}}{1986}]{couchman:86}
{Couchman} H.~M.~P.,  {Rees} M.~J.,  1986, \mnras, 221, 53

\bibitem[\protect\citeauthoryear{{Dekel} \& {Rees}}{{Dekel} \&
  {Rees}}{1987}]{dekel:87}
{Dekel} A.,  {Rees} M.~J.,  1987, Nature, 326, 455

\bibitem[\protect\citeauthoryear{{Dekel} \& {Silk}}{{Dekel} \&
  {Silk}}{1986}]{dekel:86}
{Dekel} A.,  {Silk} J.,  1986, \apj, 303, 39

\bibitem[\protect\citeauthoryear{{Efstathiou}}{{Efstathiou}}{1992}]{efstathiou%
:92}
{Efstathiou} G.,  1992, \mnras, 256, 43

\bibitem[\protect\citeauthoryear{{Gnedin}}{{Gnedin}}{2000}]{gnedin:00}
{Gnedin} N.~Y.,  2000, \apj, 542, 535

\bibitem[\protect\citeauthoryear{{Gnedin} \& {Hui}}{{Gnedin} \&
  {Hui}}{1998}]{gnedin:98}
{Gnedin} N.~Y.,  {Hui} L.,  1998, \mnras, 296, 44

\bibitem[\protect\citeauthoryear{{Goldberg}, {Jones}, {Hoyle}, {Rojas},
  {Vogeley} \& {Blanton}}{{Goldberg} et~al.}{2005}]{goldberg:05}
{Goldberg} D.~M.,  {Jones} T.~D.,  {Hoyle} F.,  {Rojas} R.~R.,  {Vogeley}
  M.~S.,    {Blanton} M.~R.,  2005, \apj, 621, 643

\bibitem[\protect\citeauthoryear{{Gottl{\" o}ber}, {{\L}okas}, {Klypin} \&
  {Hoffman}}{{Gottl{\" o}ber} et~al.}{2003}]{gottloeber:03}
{Gottl{\" o}ber} S.,  {{\L}okas} E.~L.,  {Klypin} A.,    {Hoffman} Y.,  2003,
  \mnras, 344, 715

\bibitem[\protect\citeauthoryear{{Grebel} \& {Gallagher}}{{Grebel} \&
  {Gallagher}}{2004}]{grebel:04}
{Grebel} E.~K.,  {Gallagher} J.~S.,  2004, \apjl, 610, L89

\bibitem[\protect\citeauthoryear{{Gregory} \& {Thompson}}{{Gregory} \&
  {Thompson}}{1978}]{gregory:78}
{Gregory} S.~A.,  {Thompson} L.~A.,  1978, \apj, 222, 784

\bibitem[\protect\citeauthoryear{{Grogin} \& {Geller}}{{Grogin} \&
  {Geller}}{1999}]{grogin:99}
{Grogin} N.~A.,  {Geller} M.~J.,  1999, \aj, 118, 2561

\bibitem[\protect\citeauthoryear{{Haardt} \& {Madau}}{{Haardt} \&
  {Madau}}{1996}]{haardt:96}
{Haardt} F.,  {Madau} P.,  1996, \apj, 461, 20

\bibitem[\protect\citeauthoryear{{Hoffman} \& {Shaham}}{{Hoffman} \&
  {Shaham}}{1982}]{hoffman:82}
{Hoffman} Y.,  {Shaham} J.,  1982, \apjl, 262, L23

\bibitem[\protect\citeauthoryear{{Joeveer}, {Einasto} \& {Tago}}{{Joeveer}
  et~al.}{1978}]{joeveer:78}
{Joeveer} M.,  {Einasto} J.,    {Tago} E.,  1978, \mnras, 185, 357

\bibitem[\protect\citeauthoryear{{Katz}, {Weinberg} \& {Hernquist}}{{Katz}
  et~al.}{1996}]{katz:96}
{Katz} N.,  {Weinberg} D.~H.,    {Hernquist} L.,  1996, \apjs, 105, 19

\bibitem[\protect\citeauthoryear{{Kere{\v s}}, {Katz}, {Weinberg} \&
  {Dav{\'e}}}{{Kere{\v s}} et~al.}{2005}]{keres:04}
{Kere{\v s}} D.,  {Katz} N.,  {Weinberg} D.~H.,    {Dav{\'e}} R.,  2005,
  \mnras, 363, 2

\bibitem[\protect\citeauthoryear{{Kirshner}, {Oemler}, {Schechter} \&
  {Shectman}}{{Kirshner} et~al.}{1981}]{kirshner:81}
{Kirshner} R.~P.,  {Oemler} A.,  {Schechter} P.~L.,    {Shectman} S.~A.,  1981,
  \apjl, 248, L57

\bibitem[\protect\citeauthoryear{{Klypin}, {Gottl{\" o}ber}, {Kravtsov} \&
  {Khokhlov}}{{Klypin} et~al.}{1999}]{klypin:99}
{Klypin} A.,  {Gottl{\" o}ber} S.,  {Kravtsov} A.~V.,    {Khokhlov} A.~M.,
  1999, \apj, 516, 530

\bibitem[\protect\citeauthoryear{{Klypin}, {Kravtsov}, {Bullock} \&
  {Primack}}{{Klypin} et~al.}{2001}]{klypin:01}
{Klypin} A.,  {Kravtsov} A.~V.,  {Bullock} J.~S.,    {Primack} J.~R.,  2001,
  \apj, 554, 903

\bibitem[\protect\citeauthoryear{{Klypin}, {Kravtsov}, {Valenzuela} \&
  {Prada}}{{Klypin} et~al.}{1999}]{klypin:99b}
{Klypin} A.,  {Kravtsov} A.~V.,  {Valenzuela} O.,    {Prada} F.,  1999, \apj,
  522, 82

\bibitem[\protect\citeauthoryear{{Kuhn}, {Hopp} \& {Elsaesser}}{{Kuhn}
  et~al.}{1997}]{kuhn:97}
{Kuhn} B.,  {Hopp} U.,    {Elsaesser} H.,  1997, \aap, 318, 405

\bibitem[\protect\citeauthoryear{{Lindner}, {Einasto}, {Einasto}, {Freudling},
  {Fricke}, {Lipovetsky}, {Pustilnik}, {Izotov} \& {Richter}}{{Lindner}
  et~al.}{1996}]{lindner:96}
{Lindner} U.,  {Einasto} M.,  {Einasto} J.,  {Freudling} W.,  {Fricke} K.,
  {Lipovetsky} V.,  {Pustilnik} S.,  {Izotov} Y.,    {Richter} G.,  1996, \aap,
  314, 1

\bibitem[\protect\citeauthoryear{{Mac Low} \& {Ferrara}}{{Mac Low} \&
  {Ferrara}}{1999}]{maclow:99}
{Mac Low} M.,  {Ferrara} A.,  1999, \apj, 513, 142

\bibitem[\protect\citeauthoryear{{Mateo}}{{Mateo}}{1998}]{mateo:98}
{Mateo} M.~L.,  1998, \araa, 36, 435

\bibitem[\protect\citeauthoryear{{Mathis} \& {White}}{{Mathis} \&
  {White}}{2002}]{mathis:02}
{Mathis} H.,  {White} S.~D.~M.,  2002, \mnras, 337, 1193

\bibitem[\protect\citeauthoryear{{Moore}, {Ghigna}, {Governato}, {Lake},
  {Quinn}, {Stadel} \& {Tozzi}}{{Moore} et~al.}{1999}]{moore:99}
{Moore} B.,  {Ghigna} S.,  {Governato} F.,  {Lake} G.,  {Quinn} T.,  {Stadel}
  J.,    {Tozzi} P.,  1999, \apjl, 524, L19

\bibitem[\protect\citeauthoryear{{Navarro} \& {Steinmetz}}{{Navarro} \&
  {Steinmetz}}{1997}]{navarro:97}
{Navarro} J.~F.,  {Steinmetz} M.,  1997, \apj, 478, 13

\bibitem[\protect\citeauthoryear{{Navarro} \& {Steinmetz}}{{Navarro} \&
  {Steinmetz}}{2000}]{navarro:00}
{Navarro} J.~F.,  {Steinmetz} M.,  2000, \apj, 538, 477

\bibitem[\protect\citeauthoryear{{Peebles}}{{Peebles}}{1982}]{peebles:82}
{Peebles} P.~J.~E.,  1982, \apjl, 263, L1

\bibitem[\protect\citeauthoryear{{Peebles}}{{Peebles}}{2001}]{peebles:01}
{Peebles} P.~J.~E.,  2001, \apj, 557, 495

\bibitem[\protect\citeauthoryear{{Popescu}, {Hopp} \& {Elsaesser}}{{Popescu}
  et~al.}{1997}]{popescu:97}
{Popescu} C.~C.,  {Hopp} U.,    {Elsaesser} H.,  1997, \aap, 325, 881

\bibitem[\protect\citeauthoryear{{Power}, {Navarro}, {Jenkins}, {Frenk},
  {White}, {Springel}, {Stadel} \& {Quinn}}{{Power} et~al.}{2003}]{power:03}
{Power} C.,  {Navarro} J.~F.,  {Jenkins} A.,  {Frenk} C.~S.,  {White} S.~D.~M.,
   {Springel} V.,  {Stadel} J.,    {Quinn} T.,  2003, \mnras, 338, 14

\bibitem[\protect\citeauthoryear{{Rees}}{{Rees}}{1986}]{rees:86}
{Rees} M.~J.,  1986, \mnras, 218, 25P

\bibitem[\protect\citeauthoryear{{Somerville}}{{Somerville}}{2002}]{somerville%
:02}
{Somerville} R.~S.,  2002, \apj, 572, L23

\bibitem[\protect\citeauthoryear{{Spergel}, {Bean}, {Dore'} \& et
  al.}{{Spergel} et~al.}{2006}]{spergel:06}
{Spergel} D.~N.,  {Bean} R.,  {Dore'} O.,    et al. 2006, ArXiv Astrophysics
  e-prints

\bibitem[\protect\citeauthoryear{{Spergel}, {Verde}, {Peiris}, {Komatsu},
  {Nolta}, {Bennett}, {Halpern}, {Hinshaw}, {Jarosik}, {Kogut}, {Limon},
  {Meyer}, {Page}, {Tucker}, {Weiland}, {Wollack} \& {Wright}}{{Spergel}
  et~al.}{2003}]{spergel:03}
{Spergel} D.~N.,  {Verde} L.,  {Peiris} H.~V.,  {Komatsu} E.,  {Nolta} M.~R.,
  {Bennett} C.~L.,  {Halpern} M.,  {Hinshaw} G.,  {Jarosik} N.,  {Kogut} A.,
  {Limon} M.,  {Meyer} S.~S.,  {Page} L.,  {Tucker} G.~S.,  {Weiland} J.~L.,
  {Wollack} E.,    {Wright} E.~L.,  2003, \apjs, 148, 175

\bibitem[\protect\citeauthoryear{{Springel} \& {Hernquist}}{{Springel} \&
  {Hernquist}}{2002}]{springel:02}
{Springel} V.,  {Hernquist} L.,  2002, \mnras, 333, 649

\bibitem[\protect\citeauthoryear{{Springel} \& {Hernquist}}{{Springel} \&
  {Hernquist}}{2003}]{springel:03}
{Springel} V.,  {Hernquist} L.,  2003, \mnras, 339, 289

\bibitem[\protect\citeauthoryear{{Springel}, {Yoshida} \& {White}}{{Springel}
  et~al.}{2001}]{springel:01}
{Springel} V.,  {Yoshida} N.,    {White} S.~D.~M.,  2001, New Astronomy, 6, 79

\bibitem[\protect\citeauthoryear{{Susa} \& {Umemura}}{{Susa} \&
  {Umemura}}{2004a}]{susa:04}
{Susa} H.,  {Umemura} M.,  2004a, \apj, 600, 1

\bibitem[\protect\citeauthoryear{{Susa} \& {Umemura}}{{Susa} \&
  {Umemura}}{2004b}]{susa:04b}
{Susa} H.,  {Umemura} M.,  2004b, \apjl, 610, L5

\bibitem[\protect\citeauthoryear{{Tassis}, {Abel}, {Bryan} \&
  {Norman}}{{Tassis} et~al.}{2003}]{tassis:03}
{Tassis} K.,  {Abel} T.,  {Bryan} G.~L.,    {Norman} M.~L.,  2003, \apj, 587,
  13

\bibitem[\protect\citeauthoryear{{Thoul} \& {Weinberg}}{{Thoul} \&
  {Weinberg}}{1996}]{thoul:96}
{Thoul} A.~A.,  {Weinberg} D.~H.,  1996, \apj, 465, 608

\bibitem[\protect\citeauthoryear{{Umemura} \& {Ikeuchi}}{{Umemura} \&
  {Ikeuchi}}{1984}]{umemura:84}
{Umemura} M.,  {Ikeuchi} S.,  1984, Prog.~Theor.~Phys., 72, 47

\bibitem[\protect\citeauthoryear{{van de Weygaert} \& {van Kampen}}{{van de
  Weygaert} \& {van Kampen}}{1993}]{weygaert:93}
{van de Weygaert} R.,  {van Kampen} E.,  1993, \mnras, 263, 481

\bibitem[\protect\citeauthoryear{{van den Bosch}}{{van den
  Bosch}}{2002}]{bosch:02}
{van den Bosch} F.~C.,  2002, \mnras, 331, 98

\bibitem[\protect\citeauthoryear{{Wechsler}, {Bullock}, {Primack}, {Kravtsov}
  \& {Dekel}}{{Wechsler} et~al.}{2002}]{wechsler:02}
{Wechsler} R.~H.,  {Bullock} J.~S.,  {Primack} J.~R.,  {Kravtsov} A.~V.,
  {Dekel} A.,  2002, \apj, 568, 52

\bibitem[\protect\citeauthoryear{{Yepes}, {Kates}, {Khokhlov} \&
  {Klypin}}{{Yepes} et~al.}{1997}]{yepes:97}
{Yepes} G.,  {Kates} R.,  {Khokhlov} A.,    {Klypin} A.,  1997, \mnras, 284,
  235

\end{thebibliography}

\end{document}